\documentclass[11pt]{article}

\usepackage{fullpage}
\usepackage{amsmath}
\usepackage{amssymb}
\usepackage{mathrsfs}
\usepackage{setspace}
\usepackage{bbm}
\usepackage{dsfont}

\usepackage{graphics}

\usepackage[latin1]{inputenc}
\usepackage[pdftex]{graphicx}

\usepackage{color}
\usepackage[american]{babel}

\usepackage[colorlinks=true]{hyperref} 
\hypersetup{
    bookmarks=true,         
    unicode=false,          
    pdftoolbar=true,        
    pdfmenubar=true,        
    pdffitwindow=false,     
    pdfstartview={FitH},    
    pdftitle={My title},    
    pdfauthor={Author},     
    pdfsubject={Subject},   
    pdfcreator={Creator},   
    pdfproducer={Producer}, 
    pdfkeywords={keyword1} {key2} {key3}, 
    pdfnewwindow=true,      
    colorlinks=true,       
    linkcolor=blue,          
    citecolor=red,        
    filecolor=magenta,      
    urlcolor=cyan           
}

\newcommand{\be}{\begin{equation}}
\newcommand{\ee}{\end{equation}}
\newcommand{\bes}{\begin{equation*}}
\newcommand{\ees}{\end{equation*}}
\newcommand{\bea}{\begin{eqnarray}}
\newcommand{\eea}{\end{eqnarray}}
\newcommand{\beas}{\begin{eqnarray*}}
\newcommand{\eeas}{\end{eqnarray*}}

\def\Tr{{\rm Tr}}
\def\le{\left}
\def\ri{\right}

\def\CM{{\cal M}}

\def\CO{{\cal O}}

\def\CR{{\cal R}}

\newcommand{\tr}{\mbox{tr}}

\newcommand{\RR}{\mathbb{R}}
\newcommand{\PP}{\mathbb{P}}
\newcommand{\ZZ}{\mathbb{Z}}

\newcommand{\R}{\mathbb{R}}

\begin{document}
\numberwithin{equation}{section}
{
\begin{titlepage}
\begin{center}

\hfill \\
\hfill \\
\vskip 0.75in


{\Large \bf Unravelling Holographic Entanglement Entropy in  Higher Spin Theories}\\

\vskip 0.4in

 { Alejandra Castro and Eva Llabr\'es}\\

\vskip 0.3in

{\it Institute for Theoretical Physics, University of Amsterdam,
Science Park 904, Postbus 94485, 1090 GL Amsterdam, The Netherlands}\\

\vskip 0.3in

\end{center}

\vskip 0.35in

\begin{abstract} 
\noindent
There are two proposals that compute holographic  entanglement entropy in AdS$_3$ higher spin theories based on $SL(N,\R)$ Chern-Simons theory. We show explicitly that these two proposals are equivalent. We also designed two methods that solve systematically the equations for arbitrary $N$. For finite charge backgrounds in AdS$_3$, we find exact agreement between our expressions and the short interval correction of the entanglement entropy for an excited state in a CFT$_2$.  

\end{abstract}

\vfill

\noindent \today

\end{titlepage}
}

\newpage

\tableofcontents

\section{Introduction}

Entanglement entropy is an interesting probe in a holographic setup: it is a boundary observable that directly probes the local  geometrical data of the bulk gravitational theory.  Holographic entanglement entropy might describe  how geometry emerges  in a classical theory of gravity from a quantum theory.

There are, however, many gravitational theories where  the notion of local geometrical quantities is rather unnatural. Higher spin theories, as originally formulated by Vasiliev,  are such a  class of gravitational theories with non-local interactions among a generically infinite tower of fields. The enlarged gauge redundancies of the fields act nontrivially and unfamiliarly on the metric; the usual Riemannian  definitions fall short for these theories. Nevertheless, these theories do have a seemingly healthy dual description in terms of a CFT.  In this case,  entropy (either thermal or from entanglement) in the dual theory  will provide guidance: the object that captures holographic entanglement entropy in higher spin theories can give a generalized definition of geometry. One of the goals of this program is to quantify this new definition and its possible repercussions. 

Progress has been made towards this direction. For a simple class of higher spin theories based on three dimensional $SL(N,\RR)$ Chern-Simons theory, there are two proposals for what is the appropriate object that captures entanglement entropy \cite{Ammon:2013hba,deBoer:2013vca}. Both of these proposals consider a Wilson line as the correct object that replaces the notion of geodesic length. In \cite{deBoer:2013vca} one of the key ingredients was to search for a composite line operator that was invariant under local Lorentz transformations. In \cite{Ammon:2013hba} the goal was to design a Wilson line that captured the dynamics of a massive particle.  The details of each proposal are rather different, but there was some evidence that both were equivalent since the reported results for  $SL(3)$ higher spin gravity are the same at leading order in the coupling. Here we will {\it prove} why and how they agree.  

In order to construct our proof, the first step will be to generalize the proposal of \cite{Ammon:2013hba} to $SL(N,\RR)$ Chern-Simons.\footnote{Our generalization will not only include a massive representation, but a large class of unitary representations that carry  higher spin charges.} This is done in Section \ref{sec:3}. We will give an explicit construction of the Wilson line, and  in this process we will discuss how to evaluate the saddle point value of the operator. These methods depend on the representation used for the background connections. It turns out, that the composite operator constructed in \cite{deBoer:2013vca} is a clever way to get the final answer. The authors in \cite{deBoer:2013vca} were only guided by symmetries and consistency conditions, which shows that, in this particular case, simple physical requirements on the observable are enough to capture the dynamics.  

In Section \ref{sec:5}, we develop as well two methods to explicitly evaluate the Wilson line as a function of the background $SL(N,\RR)$ fields. The first method uses the fundamental representation of the algebra. It has the advantage that it gives an exact answer for any  range of the parameters, but it is somewhat tedious to extract certain features from the answer. The second method relies on a small interval expansion of the composite Wilson line defined in \cite{deBoer:2013vca}. This small interval expansion captures first correction to the relative entropy of a pure state with respect to the vacuum. The result is well known to be  universal in a CFT$_2$ \cite{Alcaraz:2011tn}, and our results match up perfectly with this universality for any value of $N$.  We end our discussion with some open questions in Section \ref{sec:dd}. 

\section{The shortest introduction to AdS$_3$ higher spin gravity}\label{sec:int}

The simplest way to craft a higher spin theory follows from the famous observation that three dimensional Einstein gravity with a negative cosmological constant can be reformulated  as a $SL(2,\RR)\times SL(2,\RR)$ Chern-Simons theory  \cite{Achucarro:1987vz,Witten:1988hc}. By simply taking instead the gauge group to be $SL(N,\RR) \times SL(N,\RR)$, we will  produce a non-trivial theory for symmetric tensors of spin $s=2,3,\ldots, N$ \cite{Blencowe:1988gj}. 

The action of the  $SL(N,\R) \times SL(N,\R)$   Chern-Simons theory is given
by
\be\label{actiona}
S=S_{CS}[A]-S_{CS}[\bar A]~,\quad S_{CS}[A]={k\over 4\pi}  \, \int_{\CM}  \tr \Big( A\wedge dA+ {2\over 3} A\wedge A\wedge A\Big)~.
\ee
Here $\CM$ is the 3-manifold that supports the $sl(N,\R)$ algebra valued connections $A$ and $\bar A$, and the trace `$\tr$' denotes the invariant quadratic form of the Lie algebra as defined in appendix \ref{app:sln}. The equations of motion following from \eqref{actiona} are
\be\label{flat}
dA+ A\wedge A =0~, \quad d\bar A+ \bar A\wedge \bar A =0~.
\ee

The metric and higher spin fields are obtained from the Chern-Simons connection as symmetric, traceless tensors that transform in the spin $s$ representation of $SL(2,\RR)$. For example, the metric and the spin three field can be expressed as follows
\be\label{metric}
g_{\mu\nu} \sim  \tr\big (e_\mu e_\nu\big)~ , \quad  \phi_{\mu \nu\rho}\sim \tr \big(e_{(\mu } e_{\nu} e_{\rho)}\big)~,
\ee
where, in line with the pure gravity case, one defines 
\be\label{vielbeina}
e={\ell \over 2} \big(A-\bar A\big)~, 
\ee
and we introduced the  AdS radius $\ell $. The metric and higher spin fields can then be expressed in terms of trace invariants of the vielbein \cite{Campoleoni:2010zq,Campoleoni:2011hg}, with the total number of inequivalent invariants being $N-1$ for $sl(N,\R)$. This definition  for metric-like fields is appropriate for the principal embedding of $sl(2,\RR)$ in $sl(N,\RR)$.\footnote{Non-principal embeddings of $sl(2,\RR)$ in $sl(N,\RR)$ give a different gravitational interpretation of the Chern-Simons theory. Each inequivalent embedding generates a different spectrum of the theory. }

 The relation between the Chern-Simons level and the gravitational couplings is
\be
k={\ell \over8 G_3 \epsilon_N}~, \quad \epsilon_N\equiv \tr_f(L_0L_0)={1\over 12} N(N^2-1)~,
\ee
in accordance with the pure gravity limit. The notation $\tr_f$ denotes a trace in the fundamental representation of $sl(N,\R)$, and $L_0$ is given in \eqref{l0prin}. The central charge of the asymptotic symmetry group is \cite{Henneaux:2010xg,Campoleoni:2010zq}
\be\label{cc}
c=12k \epsilon_N = {3\ell\over 2 G}~.
\ee

For the immediate purpose of this work this is all we need to know about AdS$_3$ higher spin gravity. For a more  complete discussion see for example \cite{Bekaert:2005vh,Gaberdiel:2012uj,Ammon:2012wc}.

\section{Wilson lines in $SL(N,\R)$ Chern-Simons}\label{sec:3}

There are currently two seemingly different proposals to compute holographic entanglement entropy in AdS$_3$ higher spin theories. The proposal of the authors in \cite{Ammon:2013hba} states that entanglement of the dual theory is captured by a massive Wilson line
\be\label{seew}
S_{\rm EE}=-\log\left({W_{\mathcal{R}}(C)}\right)~,
\ee
where $W_{\mathcal{R}}(C)$ is a bulk Wilson line defined as:
\begin{equation}\label{Wilsonfunc}
 W_{\mathcal{R}}(C)=\text{Tr}_{\mathcal{R}}\left(\mathcal{P}\,\text{exp}\int_C (A+\bar{A}) \right)~.
\end{equation}
Here $A$ and $\bar{A}$ are the connections representing a higher spin background in  $SL(N,\mathbb{R})$ Chern-Simons theory. The representation $\mathcal{R}$ is the infinite-dimensional highest-weight representation of $sl(N,\mathbb{R})$, and $C$ is a curve in the bulk that connects the end points of the interval of width $\Delta x$ in the boundary. 

The other proposal in the market to compute entanglement entropy is given by the following object \cite{deBoer:2013vca}:
\be\label{eq:jj}
S_{\rm EE}=k \,\text{log}\left[\lim_{\rho_0\to\infty}W^{\rm comp}_{\mathcal{R}_N}(C) \right]_{\rho_0=\rho_f=\rho_i}~,
\ee
where the quantity inside the logarithm is a composite Wilson line defined as
\begin{equation}\label{compWilson}
W^{\rm comp}_{\mathcal{R}_N}(C)=\text{Tr}_{\mathcal{R}_N}\left[\mathcal{P}\,\text{exp}\left(\int_C \bar{A}\right)\mathcal{P}\,\text{exp}\left(-\int_C{A}\right)\right]~,
\end{equation}
where $C$ is the same curve as in \eqref{Wilsonfunc}. The trace is taken in a finite-dimensional representation, denoted  $\mathcal{R}_N$, which is different for every $N$. 

It was noted in \cite{Ammon:2013hba,deBoer:2013vca} that for explicit backgrounds in $SL(3,\mathbb{R})$  \eqref{eq:jj} reported the same answer as the saddle point approximation of \eqref{seew}. However, the  proposals look very different! In this section we will show that, in the semiclassical limit, they are equivalent for an open boundary interval. To do so, we will first generalise the proposal of \cite{Ammon:2013hba} to $SL(N,\mathbb{R})$. In the process of finding an efficient and systematic way to evaluate \eqref{Wilsonfunc}, we will show how  the composite Wilson line \eqref{compWilson} makes its appearance. 

In this section we will only focus on higher spin theories based on the principal embedding of $SL(2,\R)$ in $SL(N,\R)$. See appendix \ref{app:de} for the generalization to non-principal embeddings. 

\subsection{Wilson line and massive particles}\label{masspart}

As anticipated, we would like the Wilson line (\ref{Wilsonfunc}) to give information about the entanglement entropy of an open interval $\Delta x$ in the CFT. In 3d  Einstein gravity, the calculation of the entanglement entropy is equivalent to computing the length of a geodesic connecting the endpoints of $\Delta x$ \cite{Ryu:2006bv,Ryu:2006ef,Hubeny:2007xt}. A geodesic can be understood as the trajectory followed by a massive point particle. Our Wilson line should mimic the dynamics of this massive particle, and hence as a minimal requirement it should be able to carry the data of this particle. 

A point particle in the classical limit is characterized by at least one continuous parameter: the mass $m$. This data is stored in the representation $\CR$ that defines the Wilson line. An infinite-dimensional representation of $sl(N,\R)\oplus sl(N,\R)$ will do the trick: it allows for continuous parameters which we can identify with a mass.\footnote{Moreover, these infinite dimensional representations can  be unitary. It can be proven that all finite-dimensional representations of $sl(N,\mathbb{R})$ are non-unitary.} In particular, we will work with the so-called highest-weight representation. 
Consider the  $sl(N,\mathbb{R})$ algebra in (\ref{algsln}), and we define the highest-weight state of the representation as $|{\rm hw}\rangle\equiv|h,w_3,...,w_N\rangle$ with the following properties:
\be
\begin{array}{ll}
L_0|{\rm hw}\rangle=h|{\rm hw}\rangle\,,& \qquad L_{1}|{\rm hw}\rangle=0\,, \\
W^{(s)}_{0}|{\rm hw}\rangle=w_s|{\rm hw}\rangle\,, & \qquad W^{(s)}_{j}|{\rm hw}\rangle=0\,,\quad j=1,\ldots, s-1\,.
\end{array}
\ee
The constants $h$ and $w_s$ with $s=3,\ldots, N$  are the parameters defining the representation. $|{\rm hw}\rangle$ is annihilated by the lowering operators; a descendant state is created by acting with the raising operators: $W^{(s)}_{-j}$ and $L_{-1}$.   With this, the Wilson line in the infinite-dimensional highest-weight representation of $sl(N,\mathbb{R})\times sl(N,\mathbb{R})$ is labelled by two towers of quantum numbers: $(h, w_s)$ and $(\bar h, \bar w_s)$. In particular the mass $\hat m$ and orbital spin $\hat s$ are given by
\be\label{eq:cms}
\ell \hat m= h+\bar h~,\quad \hat s=\bar h-h~.
\ee
For the purpose of computing entanglement entropy we would like for the representation to only carry mass and no other quantum numbers. Hence we will set
\be\label{eq:hhw0}
h= \bar h~,\quad w_s = \bar w_s =0 ~,\quad \forall s~.
\ee
We have to fix as well the value of $\hat m$ in order to make contact with entanglement; this will be done in section \ref{sec:cone}. Of course this choice of representation can be modified, but this will change the interpretation of the Wilson line in terms of the dual theory. For instance one could design probes that carry higher spin charge or orbital spin; the interpretation of this object in the dual CFT interpretation will be different, but still rather interesting. See \cite{Hijano:2014sqa} for the case when $w_3=\bar w_3 \neq 0$  in $SL(3,\R)$ higher spin theory, and  see \cite{Castro:2014tta} for a discussion when $\hat s\neq 0$.

\subsection{Path integral representation of the Wilson line}\label{sec:pi}

The more complex step is to actually evaluate the trace in \eqref{seew}. Following \cite{Ammon:2013hba}, we will  interpret $\cal R$ as  the Hilbert space of an auxiliary quantum mechanical system that lives on the Wilson line, and replace the trace over $\CR$ by a path integral. This auxiliary system is described by some field $U$, and  we will pick the dynamics of $U$ so that upon quantization the Hilbert space of the system will be precisely the desired representation $\CR$.
More concretely,
\begin{equation}\label{pathint}
 W_{\mathcal{R}}(C)=\int\mathcal{D}U e^{-S(U,A,\bar{A})_C}~,
\end{equation}
where the  action $S(U,A,\bar{A})_C$ has $SL(N,\mathbb{R})\times SL(N,\mathbb{R})$ as a local symmetry.
The auxiliary system is appropriately described by the following action:\footnote{As discussed in \cite{Ammon:2013hba}, the choice of the action $S(U,A,\bar{A})_C$ is not unique, it is just a useful trick. There are many auxiliary systems that will recover the trace over the representation in (\ref{Wilsonfunc}), giving the same result for the Wilson line only depending on $\mathcal{R}$ and $C$.} 
\be\label{actionfin}
S(U,A,\bar{A})_C=\int_C dy \left(\text{Tr}\left(PU^{-1}D_{y}U\right)+\lambda_2(y)\left(\text{Tr}(P^2)-c_2\right)+...+\lambda_N(y)\left(\text{Tr}(P^N)-c_N\right)\right)\,.
\ee
Here $P$ is the canonical momentum conjugate to $U$ that lives in the Lie algebra $sl(N,\mathbb{R})$. The variable $y$ parametrizes the curve ${C}$, and we pick $y\in [y_i,y_f]$. The trace $\text{Tr}(...)$ is a short-cut notation for the contraction using the Killing forms in (\ref{killing}):
\be
\text{Tr}(P^m)=h_{a_{1}... a_{m}}P^{a_{1}}...P^{a_{m}}~,\quad m=2,\ldots, N~,
\ee
where $P=P^aT_a$ and $T_a$ is a generator of $sl(N,\mathbb{R})$. The functions $\lambda_m(y)$ represent Lagrange multipliers which enforce constraints on $P$. The elements $c_m$ are the Casimir invariants $C_m$ (\ref{casimirs}) applied to the highest weight state, and contain the information of the highest-weight quantum numbers $h$ and $w_s$. Note that in this action we already implemented that $h=\bar h$ and $w_s =\bar w_s$, since there is only one momenta variable $P$. This will suffice for the discussion here, but the generalization is worthwhile studying \cite{Castro:2014tta}. 

The covariant derivative is defined as
\be
D_yU\equiv \frac{d}{dy}U+A_yU-U\bar{A}_y\,, \quad A_y\equiv A_{\mu}\frac{dx^{\mu}}{dy}~,\quad \bar A_y\equiv \bar A_{\mu}\frac{dx^{\mu}}{dy}~,
\ee
where $A$ and $\bar{A}$ are the connections that determine the background.  With these definitions we have achieved our first goal: the system is invariant under the local symmetries along the curve. The transformation properties of the fields are
\be\label{gaugeA}
A_{\mu}\rightarrow L(x^\mu(y))(A_{\mu}+\partial_{\mu})L^{-1}(x^\mu(y))\,,\qquad \bar{A}_{\mu}\rightarrow R^{-1}(x^\mu(y))(\bar{A}_{\mu}+\partial_{\mu})R(x^\mu(y))~,
\ee
and
\be\label{symloc}
U(s)\rightarrow L(x^\mu(y))U(s)R(x^\mu(y))~,\quad P(y)\rightarrow R^{-1}(x^\mu(y)) P(y) R(x^\mu(y))~,
\ee
with $L$ and $R$ being element of the group $SL(N,\R)$. 

The equations of motion are:
\begin{align}
&D_yP\equiv\frac{d}{dy}P+[\bar{A}_y,P]=0~,\cr 
&U^{-1}D_yU+2\lambda_2(y) P +3\lambda_3(y) P \times P +...+N\lambda_N (y) \underbrace{P \times ... \times P}_{N-1}=0~,\label{eomP}
\end{align}
plus the Casimirs constraints $\text{Tr}(P^m)=c_m$. The cross product is a short-cut notation for:
\be\label{eq:cross}
\underbrace{P \times ... \times P}_{m} \equiv h_{i_{1} ...i_{m+1}} P^{i_{1}} ...P^{i_{m}} T^{i_{m+1}}\,.
\ee

For an open curve $C$ we need to choose boundary conditions for $U(y)$ at the endpoints of the curve. In the pure gravity case, it is natural to ask that the answer is invariant under Lorentz transformations (since the geodesic length shares this property). In $SL(2,\mathbb{R})\times SL(2,\mathbb{R})$, the group elements $R$ and $L$ that parametrize the local Lorentz subgroup is:
\be\label{RL}
R=L^{-1}~.
\ee
A natural condition is to impose that $U(y_i)$ and $U(y_f)$ are invariant under a gauge transformation of the form (\ref{RL}); this will assure that $S_{\rm EE}$ is insensitive to Lorentz transformations. From \eqref{symloc}, we see that the only boundary conditions that satisfy this condition are:
\be\label{U1}
U(y_i)=U(y_f)=\mathds{1}~.
\ee
For higher spin gravity, the symmetry group is $SL(N,\mathbb{R})\times SL(N,\mathbb{R})$ and we cannot say that the Lorentz subgroup is described by (\ref{RL}); the condition \eqref{RL} is much bigger in this case! Still, we will impose (\ref{U1})  in the higher spin case since it is the natural generalization of the gravitational case.

\subsubsection{On-shell action}\label{sec:onshell}

In this subsection we will evaluate $W_{\cal R}(C)$ in saddle point approximation. To capture this piece,  we will find a practical way to compute the classical action (\ref{actionfin}) for any background connection. The derivations will be applicable for both open and closed curves, and we will keep $w_s\neq 0$ in this subsection. 

To evaluate \eqref{actionfin}, we start by eliminating the dependence of $U$ using equation (\ref{eomP}):
\bea
S_{\text{on}-\text{shell}}&=&\int_C dy\,\text{Tr}(PU^{-1}D_yU)\cr 
&=&-\int_C dy\,(2\lambda_2(y)\text{Tr}(P^2) +3\lambda_3(y) \text{Tr}(P^3) +...+N\lambda_N (y) \text{Tr}(P^N))\cr 
&=&-\int_C dy\left( 2\,c_2 \lambda_2(y) + 3\,c_3 \lambda_3(y)+...+N\,c_N \lambda_N(y)\right) 
\eea
where in the last line we used the Casimirs constraints to eliminate $P$. Recall that the curve $C$ is running from $y\in [y_i,y_f]$. It will be useful for us to define:
\be\label{eq:alm}
\Delta\alpha_m=\alpha_m(y_f)-\alpha_m(y_i)=\int^{y_f}_{y_i} dy\,\lambda_m(y)\,,
\ee
and with this simplified notation, the action becomes:
\be\label{eq:shell}
S_{\text{on}-\text{shell}}= -2\,c_2\Delta\alpha_2 - 3\,c_3 \Delta\alpha_3+...-N\,c_N \Delta\alpha_N~.
\ee

We need to determine $\Delta\alpha_m$  as a function of the connections $A$  and $\bar{A}$; we will follow the method used in \cite{Ammon:2013hba}. We start by building a solution when $A=\bar{A}=0$: this defines for us  $U_0(y)$ and $P_0(y)$ which from  \eqref{eomP} read
\be\label{emptysol}
U_0(y)=u_0e^{-2\,\alpha_2(y) P_0- 3\,\alpha_3(y)P_0 \times P_0 +...-N\,\alpha_N(y)P_0 \times ... \times P_0}\,,\qquad P_0(y)=P_0\,,
\ee
where $u_0$ is a constant matrix, and $\alpha_m(y)$ is defined in \eqref{eq:alm}. From here, building a solution with $A\neq 0$ and $\bar A\neq 0$ is rather simple. 
As consequence of the flatness condition (\ref{flat}), every connection can be expressed locally as a gauge transformation 
\be\label{connectionRL}
A_{\mu}=L(x)\partial_{\mu}L^{-1}(x)\,,\qquad \bar{A}_{\mu}= R^{-1}(x)\partial_{\mu}R(x)\,,
\ee
where the group elelemnts $L$ and $R$ will reproduce different background connections. This means that we can build any solution to \eqref{eomP} for connections \eqref{connectionRL} by simply acting with $L$ and $R$ on \eqref{emptysol}. This gives
\be\label{UP}
U(y)= L(x(y))U_0(y)R(x(y))~, \quad P(y)=R^{-1}(x(y))P_0(y)R(x(y))~.
\ee

Next, we impose the boundary condition \eqref{U1}; enforcing this condition on \eqref{UP} gives
\begin{align}
&\mathds{1}=U(y_i)=L(y_i)\left(u_0e^{-2\,\alpha_2(y_i) P_0- 3\,\alpha_3(y_i)P_0 \times P_0 +...-N\,\alpha_N(y_i)P_0 \times ... \times P_0}\right)R(y_i)~,\cr
&\mathds{1}=U(y_f)=L(y_f)\left(u_0e^{-2\,\alpha_2(y_f) P_0- 3\,\alpha_3(y_f)P_0 \times P_0 +...-N\,\alpha_N(y_f)P_0 \times ... \times P_0}\right)R(y_f)~.
\end{align}
If we combine both previous equations to eliminate $u_0$ we obtain
\be\label{PMor}
e^{\mathbb{P}}= M~, \quad M \equiv R(y_i)L(y_i)L^{-1}(y_f)R^{-1}(y_f)~,
\ee
where we define
\bea\label{P}
\mathbb{P}\equiv-2\Delta \alpha_2 P_0 -3\Delta \alpha_3 P_0 \times P_0 +...-N\Delta \alpha_N P_0 \times ... \times P_0~.
\eea
For a given $P_0$, \eqref{PMor} determines $\Delta\alpha_m$ as a function of the background $A$ and $\bar A$. Solving \eqref{PMor} is  the most difficult task we have ahead of us. 

To determine the on-shell action we note that $\Tr(\PP P_0) = S_{\rm on-shell}$. Hence using \eqref{PMor} we find  
\be\label{eq:son}
-\log W_{\mathcal{R}}(C)=S_{\rm on-shell}=\Tr(\log(M) P_0)~.
\ee
This gives a very general expression for the on-shell value of the effective action for both open and closed curves $C$. The specific choice of $P_0$ will determine the representation $\mathcal{R}$. For instance if we wanted to evaluate the Wilson line for a representation like  \eqref{eq:hhw0} we would use
\be
P_0 =  h L_0= \sqrt{c_2\over \tr_f (L_0L_0)}L_0~,
\ee
 and for a general representation with $(h, w_s)=(\bar h,\bar w_s)$ we would have
\be
P_0 = h L_0 + \sum_{s=2}^m w_s W^{(s)}_0~.
\ee

At this stage it useful to note that $M$, as defined in \eqref{PMor}, can also be written as
\be\label{eq:mwr}
M = \mathcal{P}\,\text{exp}\left(\int_C \bar{A}\right)\mathcal{P}\,\text{exp}\left(-\int_C{A}\right)~,
\ee
and hence $W^{\rm comp}_{\mathcal{R}_N}(C)=\Tr_{\CR_N} M$. However, \eqref{eq:son} is still not casted in the appropriate way to show that \eqref{seew} is equivalent to \eqref{eq:jj}.

\subsection{Lines: Entanglement Entropy}\label{sec:cone}

As advertised, we are interested in using $W_{\CR}(C)$ to evaluate entanglement entropy. With this application in mind, we will focus our attention to open intervals that are anchored at the boundary.  Furthermore, as argued in \cite{Ammon:2013hba}, we have to choose the {\it massive} representation \eqref{eq:hhw0}. This implies that $P_0\in sl(2,\R)$.  Without loss of generality, it is convenient to set
\be\label{P0norm}
P_0=\sqrt{\frac{c_2}{\text{tr}_f(L_{0} L_{0})}}L_{0}~.
\ee
With this choice several of simplifications occur. In particular, the Casimirs $c_m=0$ for $m\geq3$ due to our choice of Killing forms in \eqref{cas0}, and hence the on-shell action \eqref{eq:shell} reduces to
\be\label{eq:ss1}
S_{\text{on}-\text{shell}}= -2\,c_2\Delta\alpha_2~.
\ee 

To solve for  $\Delta \alpha_2$ we just need to decode $\mathbb{P}$ in \eqref{P}. Since \eqref{PMor} and \eqref{eq:ss1} are independent of the representation, for simplicity, we will first focus in the fundamental representation.
Using \eqref{P0norm} and the identities listed in appendix \ref{app:sln}, we can rewrite $\mathbb{P}$ as
\be\label{Ptil}
\mathbb{P}=\Delta \tilde{\alpha}_2 L_0 +\Delta \tilde{\alpha}_3 H^{3}+\ldots+\Delta \tilde{\alpha}_N  H^{N}\,,
\ee
where all the sums of the series belong to the Cartan subalgebra, and we defined 
\be\label{eq:tilde}
\Delta \tilde{\alpha}_2\equiv -2\Delta \alpha_2\sqrt{c_2\over \text{tr}_{f}(L_{0} L_{0})}\,,
\ee
and
\be
\Delta \tilde{\alpha}_m\equiv -m\Delta \alpha_m\left({c_2\over \text{tr}_{f}(L_{0} L_{0})}\right)^{\frac{m-1}{2}} \text{tr}_{f}(\underbrace{L_{0}\ldots L_{0}}_{m-1} H_{m}),\quad\text{for}\quad m>2~.
\ee

 Since the Cartan elements are diagonal in the fundamental representation, $\mathbb{P}$ is a diagonal matrix. In order to solve \eqref{PMor}, we should put both sides in the same basis. So, we will diagonalize $M$:
\be\label{PMdiag}
\exp(\lambda_{\mathbb{P}})=\lambda_M\,,
\ee
where $\lambda_{\mathbb{P}}$ and $\lambda_M$, are the eigenvalue matrices for $\mathbb{P}$ and $M$. In principle one could evaluate the eigenvalues and match both sides, and solve for $\Delta \alpha_2$. Instead, noting that $\text{tr}_f(H_mH_{m'})=0$ if $m\neq m'$, the trace of \eqref{PMdiag} with $L_0$ gives
\be\label{alpha}
\Delta \alpha_2 = -\frac{1}{2 \sqrt{c_2\cdot\text{tr}_{f}(L_{0} L_{0})}} \text{tr}_{f}(\text{log}({\lambda_{M}}) L_{0})~.
\ee
and hence, in the saddle point approximation, we have
\be\label{eq:wm}
\log W_{\CR}(C)=-S_{\text{on}-\text{shell}}= -\sqrt{c_2\over\text{tr}_{f}(L_{0} L_{0})}  \text{tr}_{f}(\log({\lambda_{M}}) L_{0})~.
\ee
This  result assumes an ordering of the eigenvalues of $M$; we will discuss about the implications of the ordering in the following subsection.

Our goal is to compute entanglement entropy and,  for that to be the case, the massive particle described by \eqref{P0norm} needs to implement the correct type of singularity in the background solution \cite{Lewkowycz:2013nqa}. This requirement determines uniquely $c_2$, and this can be done by analyzing the backreaction of $W_{\CR}$ on $(A,\bar A)$. We will skip the details here since it follows in a straight forward manner from either the arguments in \cite{Ammon:2013hba} or \cite{Hijano:2014sqa} applied for $SL(N,\R)$ theory. We find
\be
\sqrt{c_2\over \text{tr}_f(L_0 L_0)}= k (n-1)+ O(n-1)^2~,
\ee
with $n$ being the number of replicas that define Renyi entropies, and $k$ is the Chern-Simons level. Holographic entanglement entropy is then given by
\be\label{eq:ee1}
S_{\rm EE} = \lim_{n \to 1} \frac{1}{1-n} \log \Tr \rho ^n = \lim_{n \to 1} \frac{1}{1-n} \log(W_{\CR}(C)) =  k\,   \tr_{f}(\log({\lambda_{M}}) L_{0})~.
\ee
Operationally, we may then simply write the entanglement entropy as 
\be
S_{\rm EE} = -\log\le(W_{\CR}(C)\ri) ~,\label{eesl2}
\ee
and substitute 
\be\label{c2k}
\sqrt{c_2\over \text{tr}_f(L_0 L_0)}\to k~,
\ee
in the final answer. This automatically takes care of the $n$-dependence, but it should be kept in mind that the motivation is actually the reasoning in \eqref{eq:ee1}.

\subsubsection{Primary ordering}\label{sec:ordering}

In evaluating \eqref{eq:wm} there is an implicit choice of the ordering for the eigenvalues of $M$. In this subsection we want to make this choice explicit.

$P_0$ was fixed according to \eqref{P0norm}, but this information is washed away in \eqref{eq:wm}. The problem arises because potentially there are many choices of $P_0$ that give \eqref{alpha} as a valid solution of \eqref{PMor}. Actually, the different orders of $\lambda_M$ correspond to different momenta configurations with $c_2\neq 0$ and $c_m=0$, but not necessarily compatible with \eqref{P0norm}. Since we need to assure that the Wilson line does no carry higher spin charges, we will fix the order of $\lambda_M$ which is compatible with  $P_0\in sl(2,\mathbb{R})$.

To fix the ordering, lets study first \eqref{PMdiag} when the background is in the gravitational sector: $A,\,\bar{A}\in sl(2,\mathbb{R})$. In this case we have that $M\in SL(2,\R)$. Any diagonal matrix belonging to $SL(2,\mathbb{R})$ must be conjugated to $e^{L_0}$, therefore the eigenvalues of $M$ have the form:
\be\label{a}
\{\lambda^{(j)}_M\}=\{ z^{\frac{(N-1)}{2}},\, z^{\frac{(N-3)}{2}},\, \ldots \,,\,z^{-\frac{(N-3)}{2}},\,z^{-\frac{(N-1)}{2}}\}~,
\ee
in the fundamental representation of $SL(N,\mathbb{R})$ (see appendix \ref{app:sln}), and $z$ is a function of the parameters of the background connections. 

Since  $M\in SL(2,\R)$, from \eqref{PMor} we have that $\mathbb{P}\in sl(2,\R)$. But this is not enough: we need as well that  $P_0\in sl(2,\mathbb{R})$, and from \eqref{Ptil} and \eqref{PMor} it sets $\Delta\tilde{\alpha}_m =0$ for $ m>2$.  The eigenvalues of $\exp(\mathbb{P})$ are then
\be\label{eigP}
\{e^{\lambda^{(j)}_{\mathbb{P}}}\}=\{ e^{\frac{(N-1)}{2}\Delta\tilde{\alpha}_2},\, e^{\frac{(N-3)}{2}\Delta\tilde{\alpha}_2},\, \ldots \,,\,e^{-\frac{(N-3)}{2}\Delta \tilde{\alpha}_2},\,e^{-\frac{(N-1)}{2}\Delta\tilde{\alpha}_2}\}~,.\ee
Now lets compare \eqref{eigP} with \eqref{a}. Since these equations are  invariant under  $\Delta\tilde{\alpha}_2 \rightarrow -\Delta\tilde{\alpha}_2$, and  $z\rightarrow z^{-1}$, there are two possible orders to match the eigenvalues:    $e^{\Delta\tilde{\alpha}_2}=z^{\pm1}$. 
%
%
By construction, one of the orders gives $\Delta\tilde{\alpha}_2$ positive, and the other negative. Since $S_{\rm EE}\sim \Delta\tilde{\alpha}_2 $ and the entropy must be positive, we will pick the order in which  $\Delta\tilde{\alpha}_2>0$. This determines the ordering uniquely in the $SL(2,\mathbb{R})$ limit.

 When we turn on higher spin vevs in the background connections, the eigenvalues of $M$ will change giving raise to non zero $\Delta\tilde{\alpha}_m$ in $\mathbb{P}$. However, the matching of the eigenvalues will be determined by continuity with the $SL(2,\mathbb{R})$ limit. Provided a solution to  $\Delta\tilde \alpha_m$ was found from \eqref{PMdiag}, this solution must satisfy 
\be\label{prim}
\lim_{M \to SL(2,\R)}\Delta\tilde{\alpha}_2 >0 ~,\quad  \lim_{M\to SL(2,\R)} \Delta\tilde{\alpha}_m =0 ~, \quad m>2~.
\ee
This determines the ordering of eigenvalues in $\lambda_M$ which is compatible with $P_0\in sl(2,\mathbb{R})$, and it will be referred to as \textit{primary order}. This prescription will assure that the Wilson line indeed carries the quantum numbers \eqref{eq:hhw0}.

\subsubsection{The proof}\label{sec:proof}

 Finding the eigenvalues $\lambda_M$ is a tedious task. However, we just need the leading divergent pieces as the endpoints of $C$ asymptote to the boundary. The goal of this section is to find a different way to solve for $\Delta\alpha_2$  when $C$ ends on an open interval  at the boundary.\footnote{ Closed curves will be discussed in section \ref{sec:thermal}.} In this process we will be able to prove that \eqref{seew} is equivalent to \eqref{eq:jj}. 
 
 Our task will be divided in two steps:
 \begin{enumerate}
 \item To first understand the divergent properties of $M$ as $C$ approaches the boundary.
 \item Find a representation that easily projects out $\Delta\alpha_2$ from $M$.
 \end{enumerate}

To achieve our first task, we need to spell out more what it is assumed about the background connections $A$ and $\bar A$. We have that
\be\label{eq:m1}
M = R(y_i)L(y_i)L^{-1}(y_f)R^{-1}(y_f) ~.
\ee 
Since we are only interested in traces of $M$, i.e. its eigenvalues, we will conjugate $M$ by $R(y_i)$ to get 
\be
M~\to ~M= L(y_i)L^{-1}(y_f)R^{-1}(y_f)R(y_i) ~.
\ee

We are interested in connections \eqref{connectionRL} of the form
\begin{align}\label{RL0}
R(x^\mu)=\text{exp}\left(\int_0^x \bar a\right)b^{-1}(\rho)~,\quad 
L(x^{\mu})=b^{-1}(\rho)\,\text{exp}\left(-\int_0^x a\right)~,
\end{align}
where $b(\rho)\equiv\exp(\rho L_0)$, ${a}=a_tdt+a_xdx$ and $\bar{a}=\bar a_tdt+\bar a_xdx$. The limit $\rho\to \infty$ defines the boundary of the space. All the connections will as well satisfy
\be\label{eq:axc}
A_x =e^{\rho} L_1 + O(1)~,\quad \bar A_x= e^{-\rho} L_{-1}+O(1) ~,
\ee
which means that the backgrounds are asymptotically AdS$_3$ in accordance with e.g. \cite{Compere:2013nba, Bunster:2014mua}. This  guarantees that all backgrounds have a well defined $SL(2,\R)\times SL(2,\R)$ limit. Our curve satisfies boundary conditions
\be
\rho(y_f) = \rho(y_i)\equiv  \rho_0~,  \quad x(y_f)-x(y_i)\equiv \Delta x~,\quad t(y)={\rm constant}~, \label{path}
\ee
and we take $\rho_0\to \infty$. 
From \eqref{eq:m1} and \eqref{RL0} we have
\be\label{Mvacuum}
M=e^{-2 L_0\rho_0}e^{\Delta a_ {x} }e^{2 L_0\rho_0}e^{-\Delta\bar a_{ x}}\,,
\ee
where $\Delta a_x$ is the integral of $a_x$ with boundary conditions \eqref{path}, and an analogous definition for $\Delta \bar a_x$. It is rather clear here that divergent piece as $\rho_0\to\infty$ is governed by $L_0$. If we solve for \eqref{PMdiag} while enforcing the primary ordering of eigenvalues, we find that the leading order solution in $e^{\rho_0}$ is
\be\label{rhospin}
e^{\Delta\tilde{\alpha}_2}\sim e^{4 \rho_0},\qquad\text{and}\qquad e^{\Delta\tilde{\alpha}_m}\sim 1,\quad m>2\,.
\ee
In this expression we are highlighting the divergent piece: the symbol ``$\sim$'' denotes equality up to a non-zero function of $(\Delta a_x, \Delta \bar a_x )$. The derivation of \eqref{rhospin} is presented in appendix \ref{app:proof}.
 
 Now we move on to our second task.  We consider again equation (\ref{PMdiag}) but this time we will take its trace:
 \be\label{PR}
\text{Tr}_{\CR}(e^{{\mathbb{P}}})=\text{Tr}_{\CR}(M)\,,
\ee
where $\CR$ is a representation that we have not fixed yet. Actually, the goal is to find a representation for which it is easy to read off  $\Delta\tilde{\alpha}_2$ from equation (\ref{PR}). 

In a general representation, $\mathbb{P}$ is not diagonal anymore. However, since all its elements are Cartans, we can write:
\be\label{Ptr}
\text{Tr}_{\mathcal{R}}(e^{{\mathbb{P}}})=\sum \limits_{j} \left(e^{\Delta \tilde{\alpha}_2 }\right)^{n^{(j)}_{\mathcal{R}}}\left(e^{\Delta \tilde{\alpha}_3 }\right)^{m^{(j)}_{\mathcal{R}}}\ldots\left(e^{\Delta \tilde{\alpha}_N }\right)^{k^{(j)}_{\mathcal{R}}}\,,
\ee
where the index $j=1,\ldots,\dim({\mathcal{R}})$, and the powers $(n^{(j)}_{\mathcal{R}}$, $m^{(j)}_{\mathcal{R}}$,  \ldots, $k^{(j)}_{\mathcal{R}})$ are the eigenvalues of the Cartan elements in ${\mathbb{P}}$, which depend on the representation $\mathcal{R}$. 

We want to take $\rho_0\to\infty$  in (\ref{Ptr}); from \eqref{rhospin} the dominant term is the one with the biggest power of $e^{\Delta \tilde{\alpha}_2}$. In the following we will find a representation $\mathcal{R}$ for which the dominant term has null powers of $e^{\Delta \tilde{\alpha}_m }$ for $m>2$. For this  representation,   the limit $\rho_0\to\infty$ of $\text{Tr}_{\mathcal{R}}(M)$ will depend only on ${\Delta \tilde{\alpha}_2}$.

In order to find the suitable representation we need a way to generally characterize the powers in (\ref{Ptr}).  Since ${\mathbb{P}}$ is an element of the Cartan subalgebra $\mathfrak{h}$, we define $\vec{p}$ as its dual element in the root space $\mathfrak{h}^*$. From equation (\ref{Ptil}), the explicit form of $\vec{p}$:
\be\label{p}
\vec p=\Delta \tilde{\alpha}_2 \vec{l}_0 +\Delta \tilde{\alpha}_3 \vec{h}_3+\ldots+\Delta \tilde{\alpha}_N  \vec{h}_N\,,
\ee
where $\vec{l}_0$, $\vec{h}_s$ are the dual elements of the Cartans $L_0$, $H_s$. We can write each diagonal element of ${\mathbb{P}}$ for a general representation $\mathcal{R}$ using its defining weights $\overrightarrow{\Lambda}_{{\mathcal{R}}}^{(j)}$:
\be
\lambda^{(j)}_{\mathbb{P}}=\langle \vec p, \overrightarrow{\Lambda}_{{\mathcal{R}}}^{(j)} \rangle\,,
\ee
where $\langle...,...\rangle$ is the inner product on the root space $\mathfrak{h}^*$, defined in Appendix \ref{sec:repth}.
Using this notation, \eqref{Ptr} reads 
\be\label{eq:trr}
\text{Tr}_{\mathcal{R}}(e^{\mathbb{P}})=\sum \limits_{j} e^{\langle \vec p, \overrightarrow{\Lambda}_{{\mathcal{R}}}^{(j)} \rangle}
=\sum \limits_{j} e^{\Delta \tilde{\alpha}_2 \langle \vec{l}_0, \overrightarrow{\Lambda}_{{\mathcal{R}}}^{(j)}\rangle+{\Delta\tilde{\alpha}_3\langle \vec{h}_3, \overrightarrow{\Lambda}_{{\mathcal{R}}}^{(j)} \rangle} +\ldots+\Delta \tilde{\alpha}_N\langle \vec{h}_N, \overrightarrow{\Lambda}_{{\mathcal{R}}}^{(j)} \rangle}\,,
\ee
where the sum runs for all the weights $\overrightarrow{\Lambda}_{{\mathcal{R}}}^{(j)}$ of the representation $\mathcal{R}$. As $\rho_0\rightarrow\infty$, the leading term is the one with the weight that maximizes $\langle \vec l_0, \overrightarrow{\Lambda}_{{\mathcal{R}}}^{(j)}\rangle$.

In the principal embedding $\vec{l}_0$ is a \textit{dominant weight}, which means that $\vec{l}_0$ has positive and integer Dynkyn labels. Moreover,
\be\label{c1}
\langle\vec{l}_0, \vec{{\alpha}}_i\rangle >0 ~,\quad i=1,\ldots,N-1\,,
\ee
where $\vec{\alpha}_i$ are the simple roots of the algebra. There exists as well a unique highest weight,$\overrightarrow{\Lambda}_{\mathcal{R}}^{hw}$, which has the biggest coefficients in the basis of the simple roots, i.e., all other weights are calculated subtracting simple roots to the highest weight:
\be\label{hw}
\overrightarrow{\Lambda}_{\mathcal{R}}^{(j)}=\overrightarrow{\Lambda}_{\mathcal{R}}^{hw}-\sum\limits_{i} n^{(j)}_i\vec{\alpha}_i\,,
\ee
where $n^{(j)}_i$ are positive integers that can be found for every weight $\overrightarrow{\Lambda}_{\mathcal{R}}^{(j)}$. With (\ref{c1}) and (\ref{hw}), we see that the maximum value of the inner product $\langle\vec{l}_0, \overrightarrow{\Lambda}_{{\mathcal{R}}}^{(j)}\rangle$ arises when $\overrightarrow{\Lambda}_{{\mathcal{R}}}^{(j)}=\overrightarrow{\Lambda}_{\mathcal{R}}^{hw}$. Consequently,  the dominant term in \eqref{eq:trr} is
\be\label{eq:trrlim}
\lim_{\rho_0\to\infty}\text{Tr}_{\mathcal{R}}(e^{\mathbb{P}})=e^{\langle \vec p,\overrightarrow{\Lambda}_{{\mathcal{R}}}^{hw} \rangle}=e^{\Delta \tilde{\alpha}_2 \langle\vec l_0, \overrightarrow{\Lambda}_{{\mathcal{R}}}^{hw}\rangle+{\Delta\tilde{\alpha}_3\langle \vec h_3, \overrightarrow{\Lambda}_{{\mathcal{R}}}^{hw} \rangle} +...+\Delta \tilde{\alpha}_N\langle \vec h_N, \overrightarrow{\Lambda}_{{\mathcal{R}}}^{hw} \rangle} ~.
\ee

Now, we are ready to pick a representation that extracts just the information of the parameter $\Delta\tilde\alpha_2$. It is obvious that we have to choose a representation $\CR_N$ whose highest weight is $\overrightarrow{\Lambda}_{\mathcal{R}_N}^{hw}=\vec{l}_0$. If we do so, the expression for $\Delta \tilde{\alpha}_2$ is 
\be
\Delta \tilde{\alpha}_2=\frac{1}{\text{tr}_f(L_0 L_0)} \text{log}\left[\lim_{\rho_0\to\infty}\text{Tr}_{\mathcal{R}_N} (M) \right]\,,
\ee
where we have used $\langle\vec l_0, \vec l_0\rangle = \tr_f (L_0 L_0)$. Using equations \eqref{eq:ss1} and \eqref{eq:tilde} we finally find 
\be\label{Sprin}
S_{EE}=k\,\text{log}\left[\lim_{\rho_0\to\infty}\text{Tr}_{\mathcal{R}_N} (M) \right]~.
\ee
The dimension of $\CR_N$ is given by the Weyl formula: 
\be
\text{dim}(\mathcal{R}_N)=\prod\limits_{\vec{\alpha}>0}\frac{\langle\overrightarrow{\Lambda}_{\mathcal{R}}^{hw}+\vec{\rho},\vec{\alpha}\rangle}{\langle\vec{\rho},\vec{\alpha}\rangle}~,
\ee
with $\vec{\rho}$ the Weyl vector and $\vec{\alpha}>0$ are positive roots.  For the principal embedding $\vec l_0=\vec \rho$ and that the number of positive roots of $SL(N,\R)$ is $N(N-1)/2$. Consequently the dimension of the representation that calculates the entanglement entropy in \eqref{Sprin} is
\be
\text{dim}(\mathcal{R}_N)=2^{\frac{N(N-1)}{2}}\,.
\ee

With formula (\ref{Sprin}), we find the leading term in $e^{\rho_0}$ (the UV cutoff) of the entanglement entropy for higher spin theory with less computational effort than using \eqref{eq:wm}. The difficult portion is to write $M$ in the representation $\CR_N$. Furthermore, formula (\ref{Sprin}) allows us to do identify proposals in \cite{Ammon:2013hba} and \cite{deBoer:2013vca}. As noted in \eqref{eq:mwr} we have 
\be\label{Mcomp}
\text{Tr}_{\mathcal{R}_N}(M)=W^{\rm comp}_{\mathcal{R}_N}(f,i)\,.
\ee
 Moreover, the representation $\mathcal{R}_N$ is exactly the same that in \eqref{eq:jj}. As a consequence, we have proven that formula (\ref{eq:jj}) captures the most divergent piece  of \eqref{seew}, and hence both proposals capture holographic entanglement for AdS$_3$ higher spin gravity. We consider (\ref{Sprin}) our most important result.

\subsection{Loops: Thermal entropy}\label{sec:thermal}

In this subsection, we will show how to find the thermal entropy for a higher spin black hole using a Wilson loop. In this case, we consider periodic boundary conditions 
\be\label{WilsonloopBC}
\rho(y_i)=\rho(y_f)\,,\quad t=0\,,\quad \Delta x= x(y_f)- x(y_i)=2\pi\ell~,
\ee
where $x$ is the spatial coordinate with periodicity $ x\sim x+2\pi\ell$. In $SL(2,\mathbb{R})\times SL(2,\mathbb{R})$, the Wilson loop in the infinite dimensional representation computes the length around the horizon, which is the thermal entropy of the black hole \cite{Ammon:2013hba}. Analogously as we did for the entanglement entropy, we will show that for the representation \eqref{eq:hhw0}, the Wilson loop in $SL(N,\mathbb{R})\times SL(N,\mathbb{R})$ will recover the thermal entropy for higher spin black holes in agreement with \cite{deBoer:2013gz}.

From Section \ref{sec:onshell} we found a general expression for the on-shell value; however this expression simplifies greatly for a closed path. 
We start by noticing that the auxiliary variables of the Wilson line require
\be\label{UloopBC}
U(y_f)=U(y_i)~, \qquad P(y_f)=P(y_i)~.
\ee
Imposing these periodic conditions for $U$ in (\ref{UP}), and we get 
\be
e^{\mathbb{P}}=u_0^{-1}\left(L^{-1}(y_f)L(y_i)\right)u_0\left(R(y_i)R^{-1}(y_f)\right)\,.
\ee
Using \eqref{RL0}, we rewrite the previous equation as:
\be\label{ePth0}
e^{\mathbb{P}}=u_0^{-1}\text{exp}\left(2\pi\ell a_{ x}\right)u_0\text{exp}\left(-2\pi\ell \bar{a}_{ x}\right)\,.
\ee
Here we are assuming that $(a_x,\bar a_x)$ are constant connections. Demanding  periodicity in $P(y)$ in equation (\ref{UP}) we obtain the following condition:
\be
\left[P_0, R^{-1}(y_f)R(y_i)\right]=0\,.
\ee
which says that $P_0$ and $\bar{a}_ x$ simultaneously diagonalize, and, therefore, the same do $\mathbb{P}$ and $\bar{a}_ x$. If we denote $V$ as the matrix of eigenvectors, and $\lambda_{ x}$ and $\lambda_{\mathbb{P}}$ represent the eigenvalues, equation (\ref{ePth0}) reduces to
\be\label{ePth1}
\text{exp}(\lambda_{\mathbb{P}})=(u_0V)^{-1}\text{exp}\left(2\pi\ell a_{ x}\right)(u_0V)\text{exp}\left(-2\pi\ell \bar{\lambda}_{ x}\right)\,.
\ee
Since the left-hand-side is diagonal, consistency of the previous equation requires to choose $u_0$ such that $u_0V$ is the matrix which diagonalizes $a_{ x}$, and the right-hand-side of (\ref{ePth1}) is diagonal as well. With this choice:
\be\label{Pth}
\text{exp}(\lambda_{\mathbb{P}})=e^{2 \pi\ell (\lambda_{ x}-\bar{\lambda}_{ x})}\,.
\ee
Analogously to Section \ref{sec:onshell}, we use $\Tr(\mathbb{P}P_0)=S_{\text{on-shell}}$ to find:
\be
-\log W_{\CR}(C)=S_{\text{on-shell}}=\tr_f\le(2 \pi\ell (\lambda_{ x}-\bar{\lambda}_{ x}) P_0\ri)~.
\ee

To compute thermal entropy we choose again $P_0$ as \eqref{P0norm}, and use \eqref{c2k}. In this case, the Wilson line computes gives
\be\label{sth}
S_{\rm th} = 2 \pi k \tr_{f}\le((\lambda_{ x}-\bar{\lambda}_{ x}) L_{0}\ri)~.
\ee
 This is the generalization to $SL(N,\mathbb{R})$ for the thermal entropy found in reference \cite{Ammon:2013hba} for $SL(3,\mathbb{R})$. With this result we have reproduced by means of our formalism the thermal entropy for higher spin black hole, proposed originally in \cite{deBoer:2013gz}. If we choose $P_0 \sim W_0^{(3)}$ we would reproduce the thermal results for spin-3 entropy  defined in \cite{Hijano:2014sqa}.

\section{Entanglement entropy on finite charge backgrounds}\label{sec:5}

In this section we will evaluate our Wilson line and obtain $S_{\rm EE}$ for any background connection that satisfies Drinfeld-Sokolov boundary conditions. The backgrounds represent finite charge solutions, and the results in this section are valid for either higher spin black holes \cite{Gutperle:2011kf,Compere:2013nba,Bunster:2014mua} or a conical defect \cite{Castro:2011iw}.

More explicitly, we will consider connections of the form \eqref{RL0}-\eqref{eq:axc}, and we will implement these boundary conditions by writing
\be\label{eq:a11}
A= b^{-1} a\, b + b^{-1} d b~,\quad \bar A= b\, \bar a\, b^{-1} + b\,db^{-1} ~,
\ee
 where $b(\rho)\equiv\exp(\rho L_0)$, ${a}=a_tdt+a_xdx$ and $\bar{a}=\bar a_tdt+\bar a_xdx$, and 
 \be\label{eq:a12}
 a_x = L_1 + \sum_{s=2}^{N} q_{(s)} W^{(s)}_{-s+1} ~,\quad  \bar a_x = L_{-1} + \sum_{s=2}^{N} \bar q_{(s)} W^{(s)}_{s-1} ~.
 \ee
In this decomposition, $(a_x,\bar a_x )$  contain the information about the higher spin charges of the solutions, which up to some normalization are $(q_{(s)},\bar q_{(s)})$. In particular the conformal weights are given by 
\be\label{hhh}
h = k \tr_f(L_{-1}L_{1}) q_{(2)}  ~,\quad \bar h = k \tr_f(L_{-1}L_{1}) \bar q_{(2)}~. 
\ee
The  components $(a_t,\bar a_t )$, which are constrained by the equation of motions, contain  the conjugate potentials to the charges.\footnote{Since we will only compute  $S_{\rm EE}$ on a spatial interval, the temporal component of the connections is not relevant in the present discussion. See \cite{Compere:2013nba, Bunster:2014mua,deBoer:2014fra} for the full expressions. } This decomposition of the connection follows from the discussion in \cite{Compere:2013nba, Bunster:2014mua,deBoer:2014fra}; note that this is different from the holomorphic decomposition used in \cite{Gutperle:2011kf}.

In the following we will develop two different methods explicitly evaluate $S_{\rm EE}$ as function of $q_{(s)}$ and the length of the interval $\Delta x$.   
The first method will elaborate on solving for the eigenvalues of $M$ and evaluating \eqref{eq:ee1}. This method gives an exact answer for any  range of $\Delta x$, but it is somewhat tedious to extract certain features from the answer. The second method relies on a small interval expansion of \eqref{Sprin}. The main appeal of this limit is that the first correction relative to the vacuum is universal \cite{Alcaraz:2011tn} which gives a direct check of the proposals in \cite{Ammon:2013hba,deBoer:2013vca} for any $N$.

\subsection{Method I}
 
Our goal is to characterize the leading divergent behavior of the eigenvalues of the matrix $M$ in \eqref{PMor} so we can evaluate \eqref{eq:ee1}. Working in the fundamental representation of $sl(N,\R)$, the characteristic polynomial is given by
 \be\label{polychar1}
 \lambda_M^N+c_1\lambda_M^{N-1}+c_2\lambda_M^{N-2}+...\,c_i\lambda_M^{N-i}...+c_{N-1}\lambda_M  +c_N=0\,.
 \ee
 The coefficients $c_i$ can be written in terms of traces of powers of the matrix $M$, see e.g. \cite{coef}. The coefficient $c_i$ is the result of the following determinant:
 \be\label{ci}
c_i=
\frac{(-1)^i}{i!}\left|
\begin{array}{ccccccc}
M_1& 1  & 0&0 & \cdots& &0\\
 M_2  &  M_1 & 2& 0 & \cdots & & 0 \\
 M_3 & M_2  & M_1 & 3 & \cdots & & 0 \\
 \vdots &                   &\ddots &\ddots&&&0\\
   M_{i-1} & M_{i-2} & & \cdots & &M_1 &i-1\\
M_i & M_{i-1} & & \cdots & & M_2 &M_1
\end{array}
\right|~,
\ee
 where $M_i\equiv\text{tr}_f(M^i)$. 
 
 Since we are interested in the behavior of the eigenvalues when $\rho_0\rightarrow \infty$, it will be useful to find the leading order term of the coefficients $c_i$. To do so, we will first determine $c_i$ as a function of the eigenvalues $\lambda^{(j)}_M$.
 The first coefficient is explicitly:
\be
c_1=-M_1=-(\lambda^{(1)}_M+\lambda^{(2)}_M+...+\lambda^{(N)}_M)\,.
\ee 
Using (\ref{eigorder}), we see that only the first eigenvalue contributes to the leading order of $c_1$:
\be
c_1\sim\lambda^{(1)}_M  \sim \varepsilon^{-2(N-1)}\,,
\ee
where $\varepsilon=e^{-\rho_0}$ is the UV cutoff. The second coefficient can be written as:
\be
c_2=\frac{1}{2}\left[(M_1)^2-M_2\right]=\lambda^{(1)}_M\lambda^{(2)}_M+...+\lambda^{(N)}_M\lambda^{(1)}_M+\lambda^{(N)}_M\lambda^{(2)}_M\,,
\ee
whose leading term is:
\be
c_2\sim\lambda^{(1)}_M\lambda^{(2)}_M\sim \varepsilon^{-2(N-1)-2(N-3)}\,.
\ee
%
%
%
If we keep analyzing (\ref{ci}) for different values of $i$, we will see that the coefficient $c_i$ will always be a sum of terms of the type:
\be
c_i\propto \sum_{j\neq k\neq...\neq l}\underbrace{\lambda^{(j)}_M\lambda^{(k)}_M...\lambda^{(l)}_M}_{i-\text{terms}}\,,
\ee
where each term in the series is a multiplication of $i$ eigenvalues, and the sum runs over all possible combinations of the $j,k,l...$, which do not repeat an eigenvalue more than once. The symbol $\propto$ means that both sides are equivalent up to a numerical factor. Using (\ref{eigorder}), the leading divergence in $\rho_0$ of $c_i$ is the term that contains the first $i$-th eigenvalues:
\be\label{ci3}
c_i\sim\lambda^{(1)}_M\lambda^{(2)}_M...\lambda^{(i)}_M~.
\ee
This determines the leading power in $\varepsilon$. Defining $m_i$ as the factor which multiplies the leading $\varepsilon$-power in $c_i$, and using \eqref{lambdam}, we have
\be\label{cimi}
c_i=(-1)^i m_i\varepsilon^{-2i(N-i)}+ \ldots
\ee
where the dots stand for subleading order terms when $\rho_0\rightarrow\infty$. It is useful to notice that $c_N=(-1)^N\text{det}(M)=(-1)^N$, and  hence $m_N=1$.
In this regime, we can rewrite  (\ref{polychar1}) as:
 \be\label{polychar}
 \lambda_M^N-\frac{m_1}{\varepsilon^{2(N-1)}}\lambda_M^{N-1}+...+\, (-1)^i\frac{m_i}{\varepsilon^{2i(N-i)}}\lambda_M^{N-i}+...+(-1)^{N-1}\frac{m_{N-1}}{\varepsilon^{2(N-1)}}\lambda_M  +(-1)^N=0\,.
 \ee
 
We can find the eigenvalues as a function of $m_i$ solving the previous equation. We just need to substitute in (\ref{polychar}) the leading term of one eigenvalue found in (\ref{eigorder}), and we will see that only two terms are dominant in the equation. Not considering the rest of the terms, we can solve for each eigenvalue, and find its expression as a function of $m_i$'s. However, we have actually already found the solution through the reasoning above. From (\ref{ci3}), we see that we can write an eigenvalue in terms of the coefficients $c_i$, and with (\ref{cimi}), we write it in terms of $m_i$:
\be\label{eigmi}
 \lambda^{(i)}_M\sim-\frac{c_{i}}{c_{i-1}}\sim\frac{m_{i}}{m_{i-1}}\varepsilon^{-2(N-(2i-1))}\,,
\ee
where $i=1,...N$, and we have defined $c_0\equiv1$ and $m_0\equiv1$. Now, we are ready to write the entanglement entropy as a function of the factors $m_i$, which are easier to compute than the exact eigenvalues. First, we write (\ref{eq:ee1}) in the following form:
\be
S_{\rm EE}=k\,\tr_f(\log(\lambda_M) L_0)=k\,\log\left((\lambda^{(1)}_M)^{\frac{N-1}{2}}...\,(\lambda^{(i)}_M)^{\frac{N-(2i-1)}{2}}...\,(\lambda^{(N)}_M)^{-\frac{N-1}{2}}\right)\,.
\ee
Substituting (\ref{eigmi}) in the previous formula, we obtain the leading term of the entanglement entropy in the UV cutoff:
\be\label{eemi}
S_{\rm EE}=k\log\left(\frac{m_1m_2....m_{N-1}}{\varepsilon^{4\text{tr}_f(L_0L_0)}}\right)\,.
\ee
To evaluate $m_i$, as a function of the background and the boundary interval, we use \eqref{cimi}. That is, we first evaluate $c_i$ using \eqref{ci} as a function of the traces  $M_i$ to leading order in $\varepsilon$, and from there we  read off $m_i$ using \eqref{cimi}.

\subsubsection{Example: $SL(3)$ excited states}
As an example, we evaluate \eqref{eemi} for excited states in $SL(3)$ Chern-Simons theory. The connections are given by
 \be\label{eq:ae}
 a_x = L_1 +q_{(2)} L_{-1} +q_{(3)} W_{-2}^{(3)}~, \quad  \bar a_x = L_{-1} +\bar q_{(2)} L_{1} +\bar q_{(3)} W_{2}^{(3)}~,
 \ee
 with 
 \be
 q_{(2)}= -{C_2(\lambda) \over t^{(2)}_1} ~,\quad q_{(3)}= {C_3 (\lambda)\over t^{(3)}_2} ~, 
 \ee
and the Casimirs are given by
 \be
  C_s (\lambda)\equiv {1 \over s} \sum_{i=1}^N (\lambda_i)^s ~,
 \ee
where $\lambda_i$ are the eigenvalues of $a_x$,  $t^{(s)}_j$ are the traces in \eqref{app:tr}, and the definitions for $\bar a_x$ are analogous. In this notation, a higher spin black hole corresponds to a solution with $\lambda_i\in \RR$ (real eigenvalues), while a conical defect is a solution with $\lambda_i \in i \ZZ_3$ (imaginary eigenvalues that exponentiate to the center of $SL(3,\mathbb{C})$). 
In this notation the conformal dimensions and spin-3 charge of the background solution (for general $N$) are
\bea\label{eq:h23}
&&h={c\over N(N^2-1)} C_2(\lambda)~,\quad \bar h={c\over N(N^2-1) } C_2(\bar \lambda)~,\cr
&&w_3=\le({c\over N(N^2-1)}\ri)^{3/2} C_3(\lambda)~,\quad \bar w_3=\le({c\over N(N^2-1) }\ri)^{3/2} C_3(\bar \lambda)~,
\eea
which follow the conventions in \cite{Castro:2011iw}. 

The relevant traces to evaluate the Wilson line are
\be
c_1= -{\rm tr}_f(M) = -\frac{m_1}{\varepsilon^{4}} + \CO(\varepsilon^{-2})~,\qquad 2 c_2= {\rm tr}_f(M)^2 - {\rm tr}_f(M^2) = \frac{2 m_2}{\varepsilon^{4}} + \CO(\varepsilon^{-2}) ~.\label{m1m2def}
\ee
From \eqref{eemi}, and setting $N=3$, we have 
\bea\label{eq:ees1}
S_{\rm EE} = 2 k\log\le(\frac{\sqrt{m_1 m_2}}{\varepsilon^4}\ri)~, \quad c= 24 k~.
\eea
The values of $m_{1,2}$ for a connection of the form \eqref{eq:ae}, as a function of its eigenvalues, are
\bea\label{condefm1m2}
m_1&=& {4 \over \prod}\le[(\lambda_1-\lambda_2)e^{\lambda_3\Delta x}+(\lambda_2-\lambda_3)e^{\lambda_1\Delta x}+(\lambda_3-\lambda_1) e^{\lambda_2\Delta x} \ri]\times\cr 
&&\quad\quad\le[(\bar\lambda_2-\bar \lambda_1)e^{-\bar \lambda_3\Delta x}+(\bar\lambda_3-\bar\lambda_2)e^{-\bar \lambda_1\Delta x}+(\bar\lambda_1-\bar\lambda_3) e^{-\bar\lambda_2\Delta x} \ri] ~,\cr
m_2&=& {4 \over \prod}\le[(\lambda_2-\lambda_1)e^{-\lambda_3\Delta x}+(\lambda_3-\lambda_2)e^{-\lambda_1\Delta x}+(\lambda_1-\lambda_3) e^{-\lambda_2\Delta x} \ri] \times\cr 
&&\quad\quad \le[(\bar\lambda_1-\bar \lambda_2)e^{\bar \lambda_3\Delta x}+(\bar\lambda_2-\bar\lambda_3)e^{\bar \lambda_1\Delta x}+(\bar\lambda_3-\bar\lambda_1) e^{\bar\lambda_2\Delta x} \ri]~,
\eea
where $\sum_i \lambda_i=0$, and 
\be
\prod \equiv \prod_{i>j}(\lambda_i-\lambda_j)(\bar \lambda_i-\bar\lambda_j)~.
 \ee
We would like to emphasize that \eqref{condefm1m2} is a different function of the background charges  for the spin-3 black hole relative to those reported in  \cite{Ammon:2013hba,deBoer:2013vca}. The reason is simple: here we used \eqref{eq:a12}, where the spatial $(a_x,\bar{a}_x)$ contain the information about the spin charges \cite{Compere:2013nba, Bunster:2014mua,deBoer:2014fra}, while in \cite{Ammon:2013hba,deBoer:2013vca} the holomorphic version of the connection was used \cite{Gutperle:2011kf}. What is interesting to note is that our results for entanglement entropy will reproduce the same answer as the holomorphic proposal in section 5 of \cite{deBoer:2013vca}. The holomorphic proposal was designed such that the Wilson line was only influenced by the portion of the connection that contained the charges explicitly (with the weakness that the composite line was not gauge covariant);  we achieved the same result using \eqref{eq:a12} with the advantage that gauge covariance is restored. 

It is interesting to evaluate the small interval expansion of \eqref{eq:ees1}. We get
\bea\label{eem1m2}
S_{\rm EE}&=&{c\over 3}\log\le({\Delta x\over \varepsilon}\ri) +{k\over 12} \le(\sum_i \lambda_i^2 +\sum_i \bar \lambda_i^2\ri)\Delta x^2 +O(\Delta x^4)\cr
&=&{c\over 3}\log\le({\Delta x\over \varepsilon}\ri) +{c\over (12)^2} \le(C_2(\lambda)+C_2(\bar\lambda)\ri)\Delta x^2 +O(\Delta x^4)\,.
\eea
 This correction to the vacuum entanglement is universal for a CFT$_2$ and it was computed \cite{Alcaraz:2011tn}: the reported result there was 
\be
S_{\rm EE, excited\, state}-S_{\rm EE, vacuum}={h+\bar h\over 6}(\Delta x)^2 + O((\Delta x)^4)~,
\ee
which agrees perfectly with \eqref{eem1m2}, since \eqref{eq:h23} implies
\be\label{ccee}
{h+\bar h\over 6}={c\over (12)^2} \le(C_2(\bar \lambda)+C_2(\lambda)\ri)~.
\ee

There is another comparison that one could make. There has been progress in evaluating entanglement entropy in CFT$_2$ with $W_3$ symmetry at finite spin-3 chemical potential \cite{Datta:2014ska,Datta:2014uxa}. Our results for EE are casted as function of the charges of the background, i.e. the eigenvalues of $(a_x,\bar a_x)$, and in order to make the comparison, we need to cast the charges as function of the potentials. This was done in \cite{Datta:2014ska,Datta:2014uxa,deBoer:2013vca} using holomorphic variables, but those results apply here as well. The agreement between the bulk and boundary computation was already noted in \cite{Datta:2014ska}.

\subsection{Method II}\label{sec:solveMfund}

 Our second method starts from \eqref{Sprin} which reads 
\be
S_{\rm EE}=k\,\log\left[\lim_{\rho_0\to\infty}\text{Tr}_{\mathcal{R}_N} (M) \right]~,
\ee
and from \eqref{eq:m1} we have
\be\label{eq:m4}
\Tr_{\CR_N}(M)= \Tr_{\CR_N}\le(\le[e^{- L_0\rho_0}e^{\Delta a_ {x} }e^{ L_0\rho_0}\ri] \le[e^{L_0\rho_0}e^{-\Delta\bar a_{ x}}e^{-L_0\rho_0}\ri]\ri)~,
\ee
where we used \eqref{RL0} and \eqref{eq:a11}. The simplicity in this formula is that we only need to evaluate one trace; the difficulty is that the representation $\CR_N$ can be rather horrible. In any case, our objective is to extract the most divergent piece in $\rho_0$ and simultaneously make a small interval expansion. 

To understand the divergent structure we first consider the vacuum configuration, i.e.
\be
a_x\to a_{\rm vac}=L_1~,\quad \bar a_x \to \bar a_{\rm vac}=L_{-1}~,
\ee
which is simply AdS$_3$ in Poincare coordinates. Then it is rather simple to show that
\bea\label{eq:uu1}
e^{- L_0\rho_0}e^{\Delta a_ {\rm vac} }e^{ L_0\rho_0}
&=&\mathds{1}+ e^{\rho_0} \Delta x L_1 + {1\over 2} e^{2\rho_0} (\Delta x)^2 L_1^2+\ldots  \cr
e^{ L_0\rho_0}e^{-\Delta \bar a_ {\rm vac} }e^{ -L_0\rho_0}
&=&\mathds{1}-e^{\rho_0} \Delta x L_{-1} + {1\over 2} e^{2\rho_0} (\Delta x)^2 L_{-1}^2+\ldots
\eea
However this series terminates at some finite power of $L_{\pm1}$, and the reason being that the matrices $e^{\pm L_0\rho_0}$ will {\it not} give an arbitrarily divergent power of $e^{\rho_0}$ as we showed in appendix   \ref{app:proof}. The largest power of $e^{\rho_0}$ is determined by the largest eigenvalue of $L_0$ which in this case is $\tr_f(L_0L_0)$; this follows from the definition of $\CR_N$ which sets $\overrightarrow{\Lambda}_{\mathcal{R}_N}^{hw}=\vec{l}_0$. Furthermore, this implies that $L_{\pm1}$ are nilpotent matrices of degree $\hat n +1$\footnote{\label{foot}Note that $\hat{n}$ is an integer number, since the product $(N)(N+1)(N-1)$ is always a multiple of 6 for $N\geq 2$.}:
\be\label{eq:uu2}
(L_{1})^{\hat n +1} =0= (L_{-1})^{\hat n +1}~,\quad  \hat{n}\equiv2\tr_f(L_0L_0)= {N(N^2-1)\over 6} ~.
\ee
 Therefore, for the  vacuum we find
\be\label{eq:Winf}
\lim_{\rho_0\to\infty}\Tr_{\CR_N} (M_{\rm vac}) = {1\over (\hat n!)^2}\Tr_{\CR_N} \left[ ( L_1)^{\hat n}  (L_{-1})^{\hat n}  \right] (\Delta x)^{2\hat n} e^{2\hat n \rho_0}~,
\ee
where we used \eqref{eq:uu1} and \eqref{eq:uu2}.  Hence
\bea
S_{\rm EE, vac}&=& k \log\le[ \lim_{\rho_0\to\infty}\Tr_{\CR_N} (M_{\rm vac})\ri] \cr
&=& {c\over 3} \log({\Delta x\over \varepsilon})~,
\eea
which is the well known universal result for the vacuum entanglement entropy in a CFT$_2$. Recall that $c$ is given by \eqref{cc} and $\varepsilon=e^{-\rho_0}$.\footnote{It is interesting to note that this derivation complements nicely the choice of representation in \cite{deBoer:2013vca}: another condition that determines $\CR_N$  is asking that the most divergent piece in the composite Wilson line scale like $(\Delta x)^{4\tr_f(L_0^2)}$ as in \eqref{eq:Winf}. This power of $\Delta x$ depends on the representation and gives the correct coefficient for the log piece in $S_{\rm EE}$. } 

For the general connections of the form \eqref{eq:a12}, the logic is rather similar. Since
\be
e^{- L_0\rho_0}W_{-s+1}^{(s)}e^{ L_0\rho_0} = e^{-(s-1)\rho_0}W_{-s+1}^{(s)} ~,\quad e^{ L_0\rho_0}W_{s-1}^{(s)}e^{ -L_0\rho_0} = e^{-(s-1)\rho_0}W_{s-1}^{(s)}~,
\ee
adding background charges does not affect the most leading power of $e^{\rho_0}$, but it will affect the coefficient in front of $e^{\hat n \rho_0}$. If we Taylor expand as in \eqref{eq:uu1}, schematically we will have
\bea\label{eq:uu3}
e^{- L_0\rho_0}e^{\Delta a_ {x} }e^{ L_0\rho_0}&\sim& \mathds{1}+ \Delta x(e^{\rho_0}  L_1 + e^{-(s-1)\rho_0}q_{(s)}W_{-s+1}^{(s)} )\cr &&+ {1\over 2}  (\Delta x)^2(e^{\rho_0}  L_1 + e^{-(s-1)\rho_0}q_{(s)}W_{-s+1}^{(s)} )^2+\ldots
\eea
 We are still interested solely on the terms which grow like $e^{\hat n\rho_0}$ in \eqref{eq:uu3}. The complication now is that this can be achieved, for example, by having additional $n$ powers of $L_1$ interlaced with $n'$ powers of $W^{(s)}_{-s+1}$ such that $n-(s-1)n'=\hat n$. But say we are only interested in the first correction in $\Delta x$ away from the vacuum. Then, by inspection of \eqref{eq:uu3}, we get that the relevant term comes from terms involving $L_1$ and $L_{-1}$ solely: 
 \bea\label{eq:t1}
e^{- L_0\rho_0}e^{\Delta a_ {x} }e^{ L_0\rho_0}={1\over \hat n!} (L_1)^{\hat n}(\Delta x)^{\hat n} e^{\hat n \rho_0} + {1\over (\hat n +2)!}(\Delta x)^{\hat n+2} q_{(2)}  {\cal T}_{\hat n +2}+O(e^{(\hat n-1)\rho_0}, (\Delta x)^{\hat n +4})~,
 \eea
 where 
 \be\label{eq:Tcal}
{\cal T}_{\hat n +2}\equiv L_1 L_{-1} (L_1)^{\hat n} +  L_1 L_{1} L_{-1}(L_1)^{\hat n-1}  + \cdots +  (L_1)^{\hat n}  L_{-1}L_1\,.
 \ee
 Basically the first correction in $\Delta x$ comes from  a term in \eqref{eq:uu3} that has $\hat n+1$ powers of $L_1$  and one power of $L_{-1}$, with the condition that $L_{-1}$ cannot sit at the edge of the string. It is useful to notice that \eqref{eq:Tcal} can be rewritten as:
 \be
 {\cal T}_{\hat n +2}=-(\hat n/6)(\hat n+2)(\hat n+1)(L_1)^{\hat n}\,.
 \ee
(See Appendix \ref{Tcal} for details). Analogously, for the barred sector we have
 \be\label{eq:t2}
 e^{- L_0\rho_0}e^{\Delta \bar a_ {x} }e^{ L_0\rho_0}={1\over \hat n!} (L_{-1})^{\hat n}(\Delta x)^{\hat n} e^{\hat n \rho_0} + {1\over (\hat n +2)!}(\Delta x)^{\hat n+2} \bar q_{(2)}  \bar{\cal T}_{\hat n +2}+O(e^{(\hat n-1)\rho_0}, (\Delta x)^{\hat n +4})~,
 \ee
where 
\be\label{Tcalbar}
\bar{\cal T}_{\hat n +2}\equiv L_{-1} L_{1} (L_{-1})^{\hat n} +  L_{-1} L_{-1} L_{1}(L_{-1})^{\hat n-1}  + \cdots +  (L_{-1})^{\hat n}  L_{1}L_{-1} ~,
 \ee
which can be as well rewritten as $\bar{\cal T}_{\hat n +2}=-(\hat n/6)(\hat n+2)(\hat n+1)(L_{-1})^{\hat n}$.
 
Using \eqref{eq:t1} and \eqref{eq:t2} in \eqref{eq:m4} we find that\footnote{It is interesting to note that there is no linear correction in $\Delta x$ to \eqref{eq:fs}. In the Chern-Simons language this comes from the Drinfeld-Sokolov decomposition, and the fact that we have no low fractional spin generators in the Lie algebra.}
\bea\label{eq:fs}
S_{\rm EE}&=&k\,\log\left[\lim_{\rho_0\to\infty}\text{Tr}_{\mathcal{R}_N} (M) \right]\cr
&=& {c\over 3} \log\le({\Delta x\over \varepsilon}\ri) - k \frac{\hat n}{6} {(q_{(2)}+\bar q_{(2)})} (\Delta x)^2 + O((\Delta x)^4)\cr 
&=& {c\over 3} \log\le({\Delta x\over \varepsilon}\ri)+ {h+\bar h\over 6}(\Delta x)^2 + O((\Delta x)^4)\,,
\eea
We can compare as well with the universal correction of the entanglement entropy for the vacuum state due to the insertion of a single primary field of weight $(h,\bar h)$ \cite{Alcaraz:2011tn}.   The results perfectly agree. This provides a non-trivial check of our method to compute holographic entanglement entropy.

\section{Discussion}\label{sec:dd}

We have explicitly constructed and evaluated a Wilson line in $SL(N,\RR)$ Chern-Simons theory with the purpose  of  computing holographic entanglement entropy in higher spin theories. We showed that the two proposals \cite{Ammon:2013hba,deBoer:2013vca} are consistent with each other. Furthermore, we checked that our results are in perfect agreement with the universal corrections computed in \cite{Alcaraz:2011tn} using CFT$_2$ techniques. This is a non-trivial test that $W_\CR(C)$ is an observable  that can generalize the notion of geometry in this class of theories. Our results, applied to $SL(3)$ higher spin gravity, are as well in agreement with the perturbative results reported in  \cite{Datta:2014ska,Datta:2014uxa} for CFT$_2$ with $W_3$ symmetry. 

We would like to end this work with some open questions and future directions:
\begin{enumerate}
\item Despite our very general results, our derivations fall short in describing entanglement when infinitely many higher spin fields are present. The simplest example of such a theory would be $hs[\lambda]\times hs[\lambda]$ Chern-Simons theory. $W_\CR(C)$ should still capture both thermal and entanglement entropy in this case. The obstruction is that both methods developed in Section \ref{sec:5} use heavily finite dimensional representations of the algebra in order to analyze \eqref{PMor}.  Evaluating a Wilson line with gauge group  $hs[\lambda]$ is not impossible, but some tricks might be  needed to apply our results to the more general case.
\item  It was noticed both in \cite{Ammon:2013hba,deBoer:2013vca} that the entanglement entropy on a  higher spin black hole violated strong sub-additivity. In both papers, the holomorphic formulation of the black hole was used. Here we used canonical description of the higher spin black hole, along the lines of \cite{Compere:2013nba, Bunster:2014mua}. What is rather interesting is that for $N=3$ our results in \eqref{condefm1m2}-\eqref{eq:ees1}  behave accordingly to the strong subadditivity bounds, i.e. EE is a monotonic function in the black hole regime.\footnote{This was noted by Jochem Knuttel for a higher spin black hole in the principal embedding theory of $SL(3)$,  and we are grateful of his observation.} It is not clear under which conditions holographic entanglement entropy should obey strong sub-additivity: higher derivative corrections  or deviations from the null energy conditions could 
violate these inequalities \cite{
Callan:2012ip,Wall:2012uf}. It is not obvious how non-local interactions tamper our expectations and why our results are so sensitive to boundary conditions. Still it would be interesting to study if the decomposition \eqref{eq:a12}  would give the desired behavior for $S_
{\rm EE}$.   
\item In \cite{Hijano:2014sqa} a new ``spin'' to  the Wilson line was given by adding higher spin charges to the representation $\CR$. This is not only a novel definition in the bulk, but a new and rather mysterious observable in the CFT.  The discussion presented here easily accommodates for this new observable, with one caveat: what is the generalization of the composite Wilson line \eqref{compWilson}? $W^{\rm comp}_{\mathcal{R}_N}(C)$ is designed to only capture entanglement entropy. Perhaps the proof in Section \ref{sec:proof} can be adjusted to instead find a composite Wilson line that gives spin-3 entanglement  \cite{Hijano:2014sqa}. 
\item One aspect that has been not studied properly in this context is entanglement entropy for multiple intervals. Homology conditions, and analogous properties of the HRT formula \cite{Hubeny:2007xt} should be tested in this context as well. Understanding the effect of junctions when several Wilson lines are present in the bulk might provide better insight to global properties of these operators and their interpretation in the CFT.  
\item It will be rather useful to have further independent derivations of entanglement in a CFT$_2$ that could corroborate our results. This could be made either  by considering the large central charge limit of theories with $W_N$ symmetry (along the lines of \cite{Headrick:2010zt,Faulkner:2013yia,Hartman:2013mia}), by using modular properties of the CFT$_2$, or by exploiting conformal perturbation theory.  Some progress has been made in conformal perturbation by \cite{Datta:2014ska,Datta:2014uxa,Long:2014oxa}.  We hope to report on related topics soon \cite{bchjk}.
\item Our discussion here is strictly classical. Quantum corrections to entanglement entropy in AdS$_3$/CFT$_2$ have been discussed in \cite{Barrella:2013wja,Faulkner:2013ana,Chen:2013dxa,Perlmutter:2013paa}. It would be interesting to see if the expectation value of the Wilson line has anything to add to this topic. 
\end{enumerate}

\section*{Acknowledgements}

We are very grateful to Nabil Iqbal, Juan I. Jottar, Jochem Knuttel and  Mukund Rangamani for discussions, and we particularly thank Martin Ammon, Nabil Iqbal and Juan I. Jottar for helpful comments on the manuscript.  This work was supported by Nederlandse Organisatie voor
Wetenschappelijk Onderzoek (NWO) via a Vidi grant.  AC was  as well  supported in part by the National Science Foundation under Grant No. PHYS-1066293 and the hospitality of the Aspen Center for Physics.

\appendix

\section{Conventions for $sl(N,\R)$ algebra}\label{app:sln}

We follow the same conventions as in \cite{Castro:2011iw}. A convenient basis for the $sl(N,\mathbb{R})$ algebra is represented by
 $\{L_0,L_{\pm1}\}$, the generators in the $sl(2,\mathbb{R})$ subalgebra, and $W^{(s)}_j$, the higher spin generators with $j=-(s-1),...(s-1)$. Their commutation relations are:
\bea\label{algsln}
[L_i, L_{i'} ]&=&(i-i')L_{i+i'}\,,\\\label{algslnsl2}
[L_i, W^{(s)}_j]&=&(i(s-1)-j)W^{(s)}_{i+j}\,.\label{algslncartan}
\eea
In this notation, $L_0$ and $W^{(s)}_0$ are elements of the Cartan subalgebra, and the rest of generators are raising and lowering operators. These commutation relations represent the principal embedding of $sl(N,\mathbb{R})$. We will often use the notation
\be\label{hh}
H_m \equiv W^{(m)}_0~,\quad m=2,\ldots, N~,
\ee
where $H_2= L_0$.

An explicit representation for the other $sl(N,\R)$ generators, which is independent of the representation, is as follows:
\be\label{wsform}
W^{(s)}_j = (-1)^{s-j-1}{(s + j -1)! \over (2 s - 2)!} \underbrace{[L_{-1},[L_{-1},\ldots,[ L_{-1}}_{s-j-1 \,{\rm terms}} ,L_1^{s-1}]\ldots ] ]~.
\ee
With this definition we have
\be\label{eq:ws}
W^{(s)}_{s-1}=(L_1)^{s-1}~,\quad W^{(s)}_{-s+1}=(L_{-1})^{s-1}~.
\ee

We write the fundamental representation of $sl(N,\R)$ as follows. The $\{L_0,L_{\pm1}\}$ generators for the principal embedding of $sl(2,\RR)$ are 
\be\label{ellone}
L_{1}=
- \left(
\begin{array}{ccccccc}
0& \cdots  & & & & &0\\
 \sqrt{N-1}  &  0 & & & \cdots & &  \\
 0 & \sqrt{2(N-2)}  &0 & & & &  \\
 \vdots &                   &\ddots &\ddots&&&\\
  &     & &\sqrt{|i(N-i)|}& 0& &\\
   & &  &  &\ddots & \ddots  &\\
0& \ldots & && & \sqrt{(N-1)}&0
\end{array}
\right)~,
\ee
\be\label{ellminus}
L_{-1}=
\left(
\begin{array}{ccccccc}
 0  &  \sqrt{N-1} & & \cdots &  && 0  \\ \vdots & 0  &\sqrt{2(N-2)} & & & &  \\ & \vdots  &\ddots &\ddots & & &  \\
  &   & & 0& \sqrt{|i(N-i)|}& &\\   &  & &  &\ddots & \ddots  &\\& & & & & 0&\sqrt{(N-1)}\\ 0& \cdots  &  & &  &&0
\end{array}
\right)~,
\ee
and
\be\label{l0prin}
L_0=\text{diag}\left(\frac{N-1}{2},\,\frac{N-3}{2},...\,\frac{N+1-2i}{2},...\, -\frac{N-3}{2},\,-\frac{N-1}{2}\right)~.
\ee

The Cartan-Killing form on $sl(N,\R)$ is given by
\bea\label{xx}
{\rm tr }_f  W^{(s)}_j W^{(r)}_{j'} = t^{(s)}_j \delta^{r,s}\delta_{j,-j'}~,
\eea
and
\bea\label{app:tr}
t^{(s)}_j = (-1)^j { (s-1)!^2 (s + j -1)! (s-j-1)! \over ( 2s -1)!(2 s-2)!} N \prod_{i = 1}^{s-1} (N^2 - i^2)~.
\eea
We always use the symbol ``${\rm tr }_f$'' to denote the trace in the fundamental representation. 

The Killing form and the Casimir invariants of $sl(N,\mathbb{R})$ are defined as follows. We can construct $N-1$ symmetric tensors which are regarded as the Killing forms of the algebra. The $m$-th order tensor is:
\be\label{killing}
 h_{a_{1}\ldots a_{m}}=\text{tr}_{f}(T_{(a_{1}}\ldots T_{a_{m})})\,,
\ee
where $m=2,\ldots,N$, and $T_a$ are all the generators of the algebra. The second order Killing form, given by \eqref{xx}, is the metric of the Lie algebra:
\be\label{killing}
\eta_{a b}=\text{tr}_{f}(T_{a}T_{b})\,.
\ee
The Lie algebra metric acts lowering and raising indexes $T^a = \eta^{ab} T_b$. We can define as well $N-1$ invariant Casimirs, which compute with all the elements of the algebra. The $m$-th order Casimir element is:
\be\label{casimirs}
 C_m= h^{a_{1}\ldots a_{m}}T_{a_{1}}\ldots T_{a_{m}}~.
\ee
It will be useful, however, to re-define Casimirs such that the following condition is met: the Casimir for $T_a\in sl(2,\R)$ vanishes for $m>2$. In particular if we pick $T_a = L_0$, this can achieved by re-defining the Killing forms such that
\be\label{cas0}
h\underbrace{_{L_{0}\ldots L_{0}}}_m=0\,,\qquad \text{if}\qquad m>2~.
\ee
This choice will assure that when we set $P\sim L_0$, i.e. we have just a massive particle with \eqref{eq:hhw0}, then $c_2\neq 0$ while $c_m=0$ otherwise.

For $j\neq 0$, we built the generators $W^{(s)}_j$ in \eqref{wsform} such that they are ladder operators, and hence do not have diagonal elements in the fundamental representation. On the contrary, $L_0$ is a diagonal matrix. As a consequence:
\be\label{killW}
h_{L_{0}\ldots L_{0} W^{(s)}_j}=\text{tr}_{f}\big(L_{(0}\ldots L_{0}W^{(s)}_{j)}\big)=0\,.
\ee

Furthermore, one can show that  the only non-null Killing forms that involve $L_0$ are
\be
h_{L_{0}\ldots L_{0} H_s }=\text{tr}_{f}\big(\underbrace{L_{0}\ldots L_{0}}_{s-1} H_s \big)~.
\ee
where $H_s$  is given by \eqref{hh}. 

\subsection{Further identities}\label{Tcal}

In this appendix we will use the above definitions of $sl(N,\RR)$ to simplify 
\be\label{eq:Tcal1} 
{\cal T}_{\hat n +2}\equiv L_1 L_{-1} (L_1)^{\hat n} +  L_1 L_{1} L_{-1}(L_1)^{\hat n-1}  + \cdots +  (L_1)^{\hat n}  L_{-1}L_1~,
 \ee
 as defined in \eqref{eq:Tcal}. We start by considering the first and last term in ${\cal T}_{\hat n +2}$:
\bea
T_1\equiv L_1 L_{-1} (L_1)^{\hat n}+  (L_1)^{\hat n}  L_{-1}L_1 &=& 2 L_0 (L_1)^{\hat n} - 2  (L_1)^{\hat n}  L_0\cr
&=&2[L_0, (L_1)^{\hat n}]\cr
&=& -2\hat n(L_1)^{\hat n}~.
\eea
In the first equality we used  $[L_1,L_{-1}]=2L_0$ to swap $L_{-1}$ with $L_{1}$, and $(L_{1})^{\hat n +1}=0$. From second to third line we used \eqref{eq:ws} and \eqref{algsln} to infer what $[L_0, (L_1)^{\hat n}]$ is. We consider now the second and the next-to-last term in \eqref{eq:Tcal1}:
\bea
T_2\equiv  L_1^2 L_{-1}(L_1)^{\hat n-1} +   (L_1)^{\hat n-1}L_{-1}L_1^2  &=& 2 (L_1L_0 (L_1)^{\hat n-1 }-(L_1)^{\hat n-1}L_0 L_1) +T_1\cr
&=&2 ((L_0L_1+L_1) (L_1)^{\hat n-1}-(L_1)^{\hat n-1 }(L_1 L_0-L_1)) +T_1\cr
&=& -2\cdot2\hat n(L_1)^{\hat n}+4(L_1)^{\hat n}~.
\eea
 In the first equality, we interchanged $L_{-1}$ with $L_{1}$, and identify two of the terms with $T_1$. From second to third, we swap $L_{0}$ with $L_{1}$. We use again $[L_0, (L_1)^{\hat n}]$ to get the fourth line. We repeat a similar procedure for $T_3$:
\bea
T_3\equiv  L_1^3 L_{-1}(L_1)^{\hat n-2} +   (L_1)^{\hat n-2}L_{-1}L_1^3  &=& 2 (L_1^2L_0 (L_1)^{\hat n-2 }-(L_1)^{\hat n-2}L_0 L_1^2) +T_2\cr
&=& -3\cdot2\hat n(L_1)^{\hat n}+3\cdot4(L_1)^{\hat n}
\eea
In ${\cal T}_{\hat n +2}$ there are $\hat{n}/2$ pair of terms of this type ($\hat n$ is always an even number as defined in \eqref{eq:uu2}). We can repeat the previous trick for every term $T_r$, where $r=1,\ldots,\hat{n}/2$. We first swap $L_{-1}$ with $L_{1}$ and identify $T_{r-1}$. In the rest of the terms we exchange $L_{0}$ with $L_{1}$, to obtain only elements proportional to $(L_1)^{\hat n}$. We will easily notice that $T_r$ will be of the form:
\be
T_r=-2r\hat n(L_1)^{\hat n}+4(1+2+...+(r/2-1))(L_1)^{\hat n}\,.
\ee
To find ${\cal T}_{\hat n +2}$, we need to sum over all $T_r$. Using little bit of algebra, we arrive to: 
\be
{\cal T}_{\hat n +2}=\sum_{r=1}^{\hat n/2}T_r=-\frac{\hat n}{6}(\hat n+2)(\hat n+1)(L_1)^{\hat n}
\ee
An analogous procedure can be repeated for \eqref{Tcalbar} to find:
\be
\bar{\cal T}_{\hat n +2}=-\frac{\hat n}{6}(\hat n+2)(\hat n+1)(L_{-1})^{\hat n}~.
\ee

\section{Representation theory of simple Lie algebras}\label{sec:repth}

All definitions and useful properties of representation theory utilized  in this work can be found in general text books (we particularly used \cite{yellowpages} and \cite{repth}). However, we would like to make a special comment about the normalization chosen for the scalar product in the Lie algebra.

We consider a general simple Lie algebra $\mathfrak{g}$ with dimension $g$, prepared in the Cartan-Weyl basis:
\bea\label{car}
&\left[H_i,H_j\right]=0\,,\\
&\left[H_i,E_{\alpha}\right]=\alpha_{(i)} E_{\alpha}\,,
\eea
where the indexes run as $i,\,j=1,...h$, and  $\alpha=1,...g-h$. The generators $H_i$ are elements of the Cartan subalgebra $\mathfrak{h}$, and $E_{\alpha}$ are ladder operators. We can associate a $h$-dimensional vector $\vec{\alpha}=(\alpha_{(1)}\,...\alpha_{(h)}$) to every element $E_{\alpha}$. These vectors $\vec{\alpha}$ are called \textit{roots}, and they belong to the dual space of the Cartan subalgebra, denoted by $\mathfrak{h}^*$.

We can define a scalar product in $\mathfrak{g}$ through the Killing form. In the basis \eqref{car}, the Killing forms will always follow:
\be\label{nonorm}
\left(H_{i},H_{j}\right)=\delta_{ij}\,,\qquad \left(H_{i},E_{\beta}\right)=0\,,\qquad \left(E_{\alpha},E_{\beta}\right)=\delta_{\alpha+\beta,0}\,,
\ee
where $\delta_{ij}$ is a Kronecker delta.
Moreover, we can define a bilinear form in $\mathfrak{h}^*$, denoted by $\langle...,...\rangle$, which is directly related to the Killing form in $\mathfrak{g}$:
\be\label{innerpr}
\langle\vec{\alpha}\,,\vec{\beta}\rangle=\left(H_{\alpha},H_{\beta}\right)=\sum_i \alpha_{(i)}\beta_{(i)}\,.
\ee

We would like to remark that in this work we have used the following convention for the Killing forms in $\mathfrak{g}$:
\be\label{norm}
\left(H_{i},H_{j}\right)=\text{tr}_f(H_i,H_j)\,,\qquad \left(H_{i},E_{\beta}\right)=\text{tr}_f(H_i,E_{\beta})\,,\qquad 
\left(E_{\alpha},E_{\beta}\right)=\text{tr}_f(E_{\alpha},E_{\beta})\,,
\ee
which obviously follows (\ref{nonorm}), but with an overall normalization constant.

\section{Details on section \ref{sec:proof}}\label{app:proof}
 
In this appendix we provide the details of deriving \eqref{rhospin} in Section \ref{sec:proof}. Following the logic used in Section \ref{sec:ordering}, we first look at the limit case  $M\in SL(2,\mathbb{R})$ and make use of the fundamental representation. From \eqref{eq:m1} and \eqref{RL0} we have
\be\label{Mvacuum}
M=e^{-2 L_0\rho_0}e^{\Delta a_ {x} }e^{2 L_0\rho_0}e^{-\Delta \bar a_{ x}}\,,
\ee
where for now $\Delta a_x$ and $\Delta \bar a_x$ live in $sl(2,\mathbb{R})$. In the limit $\rho_0\rightarrow\infty$, the eigenvalues $M$ can be approximated by its leading order term in $\varepsilon\equiv e^{-\rho_0}$:
\be\label{lambdam}
\lambda_M^{(j)} \sim f_j(\Delta x)\varepsilon^{-4n_j}\,.
\ee
The index $j$ runs from $1$ to $N$. The numbers $n_j$ are integers whose value depends only on the $b(\rho)$, and $f_j$ are continuous functions of $\Delta x$ whose behavior depends on $\Delta a_{ x}$ and $\Delta\bar{a}_{ x}$. Considering $L_0$ in the fundamental representation (\ref{l0prin}), we see from (\ref{Mvacuum}) that the maximum power of $e^{\rho_0}$ in $M$ will be $2(N-1)$, and then $\{n_j\}_{max}=(N-1)/2$. \\
\\
We know that the eigenvalues of $M\in SL(2,\mathbb{R})$ follow (\ref{a}). Consequently, the same relation will hold for their leading terms. Using the freedom to $z\rightarrow z^{-1}$, we pick $\lambda_M^{(1)}$ to retain the most negative power of $\varepsilon$. Then, $n_1$ will be the highest eigenvalue of $L_0$. For simplicity, we redefine $f_1(\Delta x)\equiv\kappa^{-2(N-1)}$. Then, we find that the leading term of the eigenvalues of $M\in SL(2,\mathbb{R})$ follows
\be\label{eigordersl2}
\{\lambda^{(j)}_M\}=\{(\kappa\,\varepsilon)^{-2(N-1)},(\kappa\,\varepsilon)^{-2(N-3)},\,...\,,(\kappa\,\varepsilon)^{-2(N-(2j-1))},\,...\,,\,(\kappa\,\varepsilon)^{2(N-3)},(\kappa\,\varepsilon)^{2(N-1)}\}\,.
\ee

 We have $\Delta\alpha_2>0$ according to \eqref{prim}, and comparing \eqref{eigordersl2} with \eqref{eigP} gives
\be\label{rhovacuum}
e^{\frac{N-1}{2}\Delta\tilde{\alpha}_2}= \lambda_M^{(1)} \sim f_1(\Delta x)\varepsilon^{-2(N-1)}\,,
\ee
This is the leading divergent behavior of $\Delta\alpha_2$ when the background connections approach the $SL(2,\R)$ limit. 

We now turn on the vevs for the higher spin charges in the connection. $M$ is still of the form \eqref{Mvacuum}, with the only difference that the elements to $\Delta a_{ x}$ and $\Delta \bar{a}_{ x}$ belong to $sl(N,\mathbb{R})$, while preserving the boundary condition \eqref{eq:axc}. Because we assume continuity with the $SL(2,\R)$ limit, when the background has higher spin particles, we can characterize $(\lambda_M)_i$ as follows:\footnote{For high values of the higher spin charges the eigenvalues might cross and suffer a discontinuous change. However, there must be a vicinity where the eigenvalues change continuously when we connect the higher spin particles. In the rest of our analysis we will consider this region.}
\be\label{lambdamspin}
(\lambda_M)_j \sim g_j(\Delta x)\varepsilon^{-4n_j}~.
\ee
Since $b(\rho)$ does not change in the presence of higher spin charges, each eigenvalue remains with the same leading power of $\varepsilon$: $n_j$ is the same that in (\ref{lambdam}). However, adding new elements to $a_{ x}$ and $\bar{a}_{ x}$  does change $f_j$ into a different function $g_j$. Consequently, connecting continuously the higher spin charges to \eqref{eigordersl2}, we can characterize the leading term of the eigenvalues of $M$ as:
\be\label{eigorder}
\{\lambda^{(j)}_M\}=\{g_1\varepsilon^{-2(N-1)},\,g_2\varepsilon^{-2(N-3)},\,...\,,\,g_j\varepsilon^{-2(N-(2j-1))},\,...\,,\,{g_{N-1}}\varepsilon^{2(N-3)},{g_N}\varepsilon^{2(N-1)}\}\,.
\ee

We assume continuity in $\Delta\tilde{\alpha}_i$, and  we know that the leading  $\rho_0$-dependence in the eigenvalues is always the same. Therefore, we infer that when we connect the higher spin elements, $\Delta\tilde{\alpha}_2$ has the same $\rho_0$ power that in (\ref{rhovacuum}) while $\Delta\tilde{\alpha}_i ,\ i>2$ remains $\rho_0$-independent:
\be
e^{\Delta\tilde{\alpha}_2}\sim e^{4\rho_0},\qquad\text{and}\qquad e^{\Delta\tilde{\alpha}_i}\sim 1,\quad i>2\,.
\ee
Although the analysis has been done for the fundamental representation, a solution to (\ref{PMdiag}) must be independent of the representation. Consequently, for any representation in the principal embedding, the leading term of $\Delta\tilde{\alpha}_m ,\ m>2$ does not depend on $\rho_0$.

\section{Non-Principal Embedding}\label{app:de}

The discussion in the main sections focused on the principal embedding of $SL(2,\RR)$ in $SL(N,\RR)$. In this appendix we will extend the results of Section \ref{sec:3} to other embeddings. Actually the discussion in subsections \ref{masspart}-\ref{sec:pi} is basically embedding independent (any modification is trivial);  only portions of subsection \ref{sec:cone} need to be revisited. 

The first modification is $L_0$:   the Cartan element in the $sl(2,\mathbb{R})$ subalgebra is not \eqref{l0prin} for non-principal embeddings. This fact affects our reasoning in subsection \ref{sec:ordering}. Equations \eqref{a} and \eqref{eigP} must be replaced by:
\be\label{a2}
\{\lambda^{(j)}_M\}=\{ z^{n_1},\, z^{n_2},\, \ldots \,,\,z^{n_{N-1}},\,z^{n_N}\}~,
\ee
and
\be\label{eigP2}
\{e^{\lambda^{(j)}_{\mathbb{P}}}\}=\{ e^{{n_1}\Delta\tilde{\alpha}_2},\, e^{{n_2}\Delta\tilde{\alpha}_2},\, \ldots \,,\,e^{{n_{N-1}}\Delta \tilde{\alpha}_2},\,e^{{n_N}\Delta\tilde{\alpha}_2}\}~,
\ee
where $n_j$ with $j=1,...,N$ are the eigenvalues of $L_0$ in the embedding of interest. These eigenvalues will always have as a symmetry: $n_j=-n_{N-(j-1)}$. This implies that the relation $e^{\Delta\tilde{\alpha}_2}=z^{\pm1}$ still holds, and a solution with $\Delta\tilde{\alpha}_2>0$ can be found for every embedding. But we should notice that, for non-principal embeddings, $L_0$ might have degenerated eigenvalues. This means that when we connect the higher spin vevs, the ordering is not uniquely determined by the $SL(2,\mathbb{R})$ limit. However, all possible matchings following condition \eqref{prim}, will give the same solution for $\Delta\tilde{\alpha}_2$ from equation \eqref{eq:wm}. This 
is because with equation \eqref{eq:wm} we extract from $\lambda_M$ only the information about the $SL(2,\mathbb{R})$ subgroup, which does not depend on how we connect the higher spin vevs. Therefore, the conclusion of the subsection \ref{sec:ordering} generalized to every embedding 
is: all orders of $\lambda_M$ that accomplish the condition \eqref{prim}, give a unique solution for $\Delta\tilde{\alpha}_2$ from \eqref{eq:wm} that is compatible with $P_0\in sl(2,\mathbb{R})$.

For subsection \ref{sec:proof} the discussion is embedding independent until \eqref{eq:trr}. However we would like to comment on equation \eqref{rhospin}, which holds for any embedding, but it was deduced in Appendix \ref{app:proof} specifically for the principal. To prove that this equation is true in general, we must notice that $\lambda_M$ can be as well characterized by \eqref{lambdam} for any embedding. However, $\{n_j\}_{max}$ is now the maximum eigenvalue of $L_0$ in the embedding we are interested in. We choose $\{n_j\}_{max}\equiv n_1$. Using \eqref{eigP2}, and imposing $\Delta\tilde{\alpha}_2>0$, we find the analogous to \eqref{rhovacuum} for non-principal embeddings:
\be\label{dd:1}
e^{n_1\Delta\tilde{\alpha}_2}= \lambda_M^{(1)} \sim f_1(\Delta x)\varepsilon^{-4n_1}\,,
\ee
Turning on the higher spin vevs, and assuming continuity in $\lambda_M$ and $\Delta\tilde{\alpha}_i$, we see conclude that equation \eqref{rhospin} holds for any embedding.

Following with the analysis of subsection \ref{sec:proof},  equation \eqref{c1} does not hold for non-principal embeddings since $\vec{l}_0$ is not dominant. In this case, the maximum value of $\langle \vec{l}_0, \overrightarrow{\Lambda}_{\mathcal{R}}^{(j)}\rangle$ is not given when $\overrightarrow{\Lambda}_{\mathcal{R}}^{hw}=\vec{l}_0$, because now $\vec{l}_0$ has negoative coefficients. To solve this problem, we transform $\vec{l}_0$ to a basis of simple roots where its coefficients are positive.  Any weight can be brought to the fundamental chamber by a unique operation of the Weyl group: we define $\omega$ as the Weyl reflection that brings $\vec{l}_0$ to the basis of simple roots where it is dominant:
\be\label{weyll00}
\langle\omega(\vec{l}_0), \vec{\alpha}_i\rangle \geq 0 \qquad i=1,...,N-1\,.
\ee
%
 We are ready to infer which term will be leading in equation (\ref{eq:trr}) when $\rho_0\rightarrow\infty$. First, we will perform the Weyl reflection $\omega$ in $\overrightarrow{p}$: \footnote{Note that, the inner product is invariant under the Weyl group:
\be
\langle\vec{l}_0, \overrightarrow{\Lambda}_{{\mathcal{R}}}^{(j)}\rangle=\langle \omega(\vec{l}_0),\omega(\overrightarrow{\Lambda}_{{\mathcal{R}}}^{(j)})\rangle=\langle \omega(\vec{l}_0),\overrightarrow{\Lambda}_{{\mathcal{R}}}^{(k)}\rangle\,.
\ee
In the last equality we used that the Weyl group reshuffles the weights, and we are allowed to relabel $\omega(\overrightarrow{\Lambda}_{{\mathcal{R}}}^{(j)})=\overrightarrow{\Lambda}_{{\mathcal{R}}}^{(k)}$ (the index $k$ is not necessarily equal to $j$).}
\be
\text{Tr}_{\mathcal{R}}(e^{\mathbb{P}})=\sum\limits_{j} e^{\langle \omega(\vec{p}), \,\omega(\overrightarrow{\Lambda}_{{\mathcal{R}}}^{(j)}) \rangle}=\sum\limits_{k} e^{\langle \omega(\vec{p}), \overrightarrow{\Lambda}_{{\mathcal{R}}}^{(k)} \rangle}\,.
\ee
In the limit $\rho_0\rightarrow\infty$ we only need the maximum value of the inner product.
 With (\ref{weyll00}) and (\ref{hw}), we find that  $\langle\omega(\vec{l}_0), \overrightarrow{\Lambda}_{{\mathcal{R}}}^{(k)}\rangle$ is maximized  when  $\overrightarrow{\Lambda}_{\mathcal{R}}^{(k)}=\overrightarrow{\Lambda}_{\mathcal{R}}^{hw}$, only if $\vec{l}_0$ is in the fundamental chamber.
 Using  \eqref{dd:1}, we know that the leading term of the sum has the inner product $\langle \omega(\vec{l}_0), \overrightarrow{\Lambda}_{\mathcal{R}}^{hw}\rangle$:
\be\label{PR2}
\lim_{\rho_0\to\infty}\text{Tr}_{\mathcal{R}}(e^{\mathbb{P}})=e^{\langle \omega(\vec{p}),\overrightarrow{\Lambda}_{{\mathcal{R}}}^{hw} \rangle}=e^{\Delta \tilde{\alpha}_2 \langle\omega(\vec l_0), \overrightarrow{\Lambda}_{{\mathcal{R}}}^{hw}\rangle+{\Delta\tilde{\alpha}_3\langle \omega(\vec w_0), \overrightarrow{\Lambda}_{{\mathcal{R}}}^{hw} \rangle} +...+\Delta \tilde{\alpha}_N\langle \omega(\vec h_a), \overrightarrow{\Lambda}_{{\mathcal{R}}}^{hw} \rangle}\,.
\ee
 With an analogous reasoning to the one after equation (\ref{eq:trrlim}), we choose the representation whose highest weight is $\overrightarrow{\Lambda}_{\mathcal{R}}^{hw}\propto w(\vec l_0)$ to extract the parameter $\Delta\alpha_2$ from equation (\ref{PR2}). However, we have to notice that all weights of a representation have integer Dynkyn labels, and the same does not holds in general for $\vec l_0$. Actually, it can be shown that only when the spectrum of particles contains semi-integer spins, $\vec{l}_0$ will have semi-integer Dynkyn labels \cite{deBoer:2013vca}. We can easily solve this problem picking the highest weight as $\overrightarrow{\Lambda}_{\mathcal{R}}^{hw}=\sigma_ {1/2}\, w(\vec l_0)$, where $\sigma_{1/2}=2$ when we have semi-integer spins in the spectrum, and $\sigma_{1/2}=1$ otherwise. Then, choosing a representation $\mathcal{R}$ with  $\overrightarrow{\Lambda}_{\mathcal{R}}^{hw}=\sigma_ {1/2}\, w(\vec{l}_0)$ we find:
\be
\Delta \tilde{\alpha}_2=\frac{1}{\sigma_{1/2} \text{tr}_f(L_0 L_0)} \text{log}\left[\lim_{\rho_0\to\infty}\text{Tr}_{\mathcal{R}} (M) \right]\,,
\ee
where we have used  $\langle\omega(\vec l_0),\omega(\vec l_0)\rangle = \tr_f (L_0 L_0)$. With (\ref{eq:ss1}) and (\ref{eq:tilde}) we arrive to the equation for the entanglement entropy:
\be\label{Sother}
S_{\rm EE}=\frac{k}{\sigma_{1/2}}\text{log}\left[\lim_{\rho_0\to\infty}\text{Tr}_{\mathcal{R}} (M) \right]\,,
\ee
where $\mathcal{R}$, as explained before, must be the representation whose highest weight satisfies $\overrightarrow{\Lambda}_{\mathcal{R}}^{hw}=\sigma_ {1/2}\, w(\vec l_0)$. We must notice that this equation includes as well the result (\ref{Sprin}) for the principal embedding, where $\sigma_ {1/2}=1$ and $\vec l_0$ is already dominant. Remembering \eqref{Mcomp}, we can see that equation \eqref{Sother} is equivalent to the entanglement entropy for a general embedding proposed in \cite{deBoer:2013vca}. As a conclusion of the generalization of subsection \ref{sec:proof}, we have proven that formula \eqref{seew} captures both proposals for the entanglement entropy in higher spin gravity for any embedding.

As a final comment, we  would like to add that subsection \ref{sec:thermal} is embedding independent. Moreover, the method to find the leading divergence of the entanglement entropy developed in subsection \ref{sec:solveMfund} is easily generalizable from the logic presented in the main text.


\end{document}